\documentclass[pdflatex,sn-mathphys-num]{sn-jnl}

\usepackage{graphicx}%
\usepackage{multirow}%
\usepackage{amsmath,amssymb,amsfonts}%
\usepackage{amsthm}%
\usepackage{mathrsfs}%
\usepackage[title]{appendix}%
\usepackage{xcolor}%
\usepackage{textcomp}%
\usepackage{manyfoot}%
\usepackage{booktabs}%
\usepackage{algorithm}%
\usepackage{algorithmicx}%
\usepackage{algpseudocode}%
\usepackage{listings}%

\usepackage[T1]{fontenc}
\usepackage{xcolor}
\usepackage{hyperref}
\usepackage{booktabs}
\usepackage{multirow}
\usepackage{amssymb}
\usepackage{soul}
\usepackage{tabularx}
\usepackage{textcomp}
\usepackage{makecell}
\newcommand\JD[1]{\textcolor{black}{#1}}

\newcommand\JAC[1]{\textcolor{black}{#1}}
\usepackage{fontawesome5}
\usepackage{mdframed}
\usepackage{refcount}

\makeatletter
\providecommand{\cline}[1]{\@cline#1\@nil}
\makeatother

\mdfdefinestyle{stebox2}{
	backgroundcolor=gray!10,
	linecolor=gray!50,
	roundcorner=5pt,
	leftmargin=0pt,
	rightmargin=0pt,
	skipabove=3pt,
	skipbelow=3pt
}

\newcommand{\rqanswerfirst}[1]{%
	\begin{mdframed}[style=stebox2]
		\textbf{{Answer to RQ1:}} #1
	\end{mdframed}
}

\newcommand{\rqanswersec}[1]{%
	\begin{mdframed}[style=stebox2]
		\textbf{{Answer to RQ2:}} #1
	\end{mdframed}
}

\newcommand{\rqanswerthree}[1]{%
	\begin{mdframed}[style=stebox2]
		\textbf{{Answer to RQ3:}} #1
	\end{mdframed}
}

\newcommand{\rqanswerfour}[1]{%
	\begin{mdframed}[style=stebox2]
		\textbf{{Answer to RQ4:}} #1
	\end{mdframed}
}

\newcommand{\rqanswerfive}[1]{%
	\begin{mdframed}[style=stebox2]
		\textbf{{Answer to RQ5:}} #1
	\end{mdframed}
}

\theoremstyle{thmstyleone}%
\theoremstyle{thmstyletwo}%

\theoremstyle{thmstylethree}%

\begin{document}

\title[Article Title]{Large Language Models for Requirements Engineering: A Cross-Task Empirical Evaluation}

\author*[1]{\fnm{Jacek} \sur{D\k{a}browski}}
\email{jacek.dabrowski@lero.ie}

\author[1]{\fnm{Manjeshwar Aniruddh} \sur{Mallya}}
\email{mallyaaniruddh@gmail.com}

\author[2,3]{\fnm{Alessio} \sur{Ferrari}}
\email{alessio.ferrari@cnr.it}

\author[4]{\fnm{Mohammad Amin} \sur{Zadenoori}}
\email{amin.zadenoori@unipd.it}

\author[5]{\fnm{Yijun} \sur{Yu}}
\email{y.yu@open.ac.uk}

\affil[1]{%
	\orgname{Lero, the Research Ireland Centre for Software, University of Limerick},
	\country{Ireland}}

\affil[2]{%
	\orgname{Trinity College Dublin (TCD)},
	\country{Ireland}}

\affil[3]{%
	\orgname{Consiglio Nazionale delle Ricerche (CNR)},
	\country{Italy}}

\affil[4]{%
	\orgname{University of Padova},
	\country{Italy}}

\affil[5]{%
	\orgname{The Open University},
	\country{UK}}



\abstract{Requirements-related information is distributed across heterogeneous artefacts, including online user feedback, developer discussions, and software repositories. Extracting actionable requirements knowledge from these artefacts is labour-intensive and difficult to scale due to their volume, diversity, and unstructured nature. Recent advances in Large Language Models (LLMs) support a broad range of Requirements Engineering (RE) activities, ranging from requirements classification and traceability identification to requirements specification and traceability explanation generation. However, existing empirical evidence on the use of LLMs in RE remains fragmented across tasks, artefact types, and evaluation settings. Consequently, broader conclusions are difficult to draw about their capabilities and limitations. In particular, studies rarely provide cross-task empirical evaluations or replication packages, limiting reproducibility and the accumulation of empirical knowledge. To address this gap, this paper presents two complementary empirical studies evaluating LLMs across five RE-related activities. The first is a controlled experiment evaluating five lightweight open-source LLMs for feedback-driven requirements classification and specification generation. The second is an exploratory industrial case study evaluating two frontier LLMs for traceability link identification and traceability explanation generation using software project artefacts. Classification and traceability identification were evaluated using quantitative metrics, while generation tasks were assessed through human evaluation. The results show that LLM performance is strongly task-dependent, ranging from moderate to high across the evaluated RE activities. No single model consistently outperformed others across all tasks, highlighting that effective LLM adoption in RE depends on selecting models and prompting strategies according to the target RE task. Our contributions are: (i) the first cross-task empirical evaluation of LLMs spanning five RE-related activities, (ii) replication materials supporting transparency and reproducibility, and (iii) a broader empirical understanding of the capabilities, limitations, and practical readiness of current LLMs for RE.}



\maketitle

\section{Introduction}\label{sec:introduction}
Modern Requirements Engineering (RE) depends on extracting actionable knowledge from heterogeneous software project artefacts, including online user feedback, issue tickets, developer discussions, and repository documentation~\cite{zadenoori2025largelanguagemodelsllms, Arora2024, Ferrari2025}. These artefacts provide complementary evidence about stakeholder needs, implementation decisions, and software evolution~\cite{11190353,zadenoori2025largelanguagemodelsllms, 10.1007/s42979-020-00416-4}. This information supports core RE activities, from requirements elicitation~\cite{10.1007/s42979-020-00416-4} and classification~\cite{alhoshan2025effectivegenerativelargelanguage} to specification and traceability~\cite{Dabrowski2022a, Dabrowski2022}.

Extracting actionable requirements knowledge from these artefacts is however increasingly challenging due to their volume, diversity, and unstructured nature~\cite{8474507, DBLP:journals/software/MaalejNJR16, 10.1007/s42979-020-00416-4}. As software projects evolve, practitioners must continuously analyse information distributed across multiple artefacts to identify, classify, document, and trace requirements~\cite{Hou2024,zadenoori2025largelanguagemodelsllms, DBLP:conf/enase/Franch20a}. Consequently, automated support for analysing requirements-related artefacts has become an important research topic in RE and Software Engineering (SE)~\cite{Hou2024, DBLP:journals/software/MaalejNJR16}.

Over the past decade, numerous Natural Language Processing (NLP) and Machine Learning (ML) techniques have been proposed to support analytical RE-related activities, including user feedback analysis, requirements classification, information extraction, and traceability recovery~\cite{DBLP:conf/icse/MaalejNR19, Dabrowski2022,8474507}. Although these approaches have demonstrated promising results, they are typically designed for individual tasks, require substantial labelled data, and often generalise poorly across projects and artefact types~\cite{10.1007/s42979-020-00416-4, Ferrari2025}. Consequently, developing and maintaining task-specific solutions remains resource-intensive and difficult to generalise across projects and artefact types~\cite{Ferrari2025}.

Recent advances in Large Language Models (LLMs) offer new opportunities to overcome these limitations~\cite{Hou2024, vogelsang2024usinglargelanguagemodels, zadenoori2025largelanguagemodelsllms}. Beyond improving analytical RE activities such as requiremetns classification~\cite{DBLP:journals/corr/abs-2510-21443} and traceability analysis~\cite{10.1007/s00766-026-00460-1}, LLMs enable generative capabilities, including requirements specification~\cite{DBLP:conf/refsq/MallyaFZD26} and explanation generation~\cite{zadenoori2025largelanguagemodelsllms, DBLP:conf/refsq/HessVFHKR26}. Owing to their strong reasoning and text generation capabilities~\cite{DBLP:journals/corr/abs-2310-03533,zhao2026survey}, LLMs have rapidly gained attention across RE/SE~\cite{DBLP:journals/corr/abs-2310-03533}. Unlike conventional NLP and ML approaches, LLMs can be adapted to diverse RE activities using natural-language prompts instead of task-specific models~\cite{,https://doi.org/10.1002/spe.70029,zadenoori2025largelanguagemodelsllms}.

Despite this growing interest, empirical evidence on the use of LLMs in RE remains fragmented across tasks, artefact types, and evaluation settings~\cite{zadenoori2025largelanguagemodelsllms, DBLP:journals/corr/abs-2310-03533, DBLP:conf/refsq/FrattiniM26}. Consequently, it is difficult to establish a broader understanding of their capabilities and limitations or determine whether findings obtained in one RE context generalise to others~\cite{Dabrowski2022, Dabrowski2022a}. Moreover, few studies provide cross-task empirical evaluations or publicly available replication materials, limiting reproducibility and the accumulation of empirical knowledge~\cite{DBLP:journals/tosem/AbualhaijaADDFFF24,zadenoori2025largelanguagemodelsllms, Dabrowski2023}.

This gap raises an important question: \textbf{how do current LLMs perform across diverse RE-related activities, and how do their capabilities and limitations vary across different RE contexts?} Addressing this question requires empirical evidence spanning multiple RE activities, artefact types, and evaluation settings.

To address this gap, we present two complementary empirical studies covering five RE-related activities. 

Mallya et al.~\cite{DBLP:conf/refsq/MallyaFZD26} previously reported the first study, which evaluated five lightweight open-source LLMs on three feedback-driven activities: non-functional requirements classification, user request classification, and requirements specification generation. This manuscript substantially extends that work by introducing a second empirical study. The second study is an exploratory industrial case study that evaluates two frontier LLMs for traceability link identification and traceability explanation generation using heterogeneous software project artefacts.

The two studies intentionally evaluate different categories of LLMs. The first study investigates whether lightweight open-source LLMs can support RE tasks with predefined outputs, including classification and requirements specification. The second study evaluates frontier LLMs on traceability analysis, which requires identifying semantic relationships across multiple software artefacts and generating evidence-grounded explanations. These studies move beyond task-specific evaluations by integrating evidence from complementary RE scenarios. This provides a broader empirical perspective on the capabilities and limitations of LLMs across requirements classification, specification, traceability, and explanation generation.

Rather than proposing a new LLM architecture or prompting strategy, this work aims to build cumulative empirical evidence on the capabilities and limitations of current LLMs for RE. By combining evidence from these complementary RE-related tasks, we identify strengths, limitations, and practical challenges that influence the effective adoption of LLMs in RE. The main contributions of this paper are as follows:

\begin{itemize}
	\item the first cross-task empirical evaluation of LLMs spanning five RE-related activities;
	\item  replication materials supporting transparency and reproducibility~\cite{supplementary_materials};
	\item a broader empirical understanding of the capabilities and limitations of current LLMs for RE.
\end{itemize}

The remainder of this paper is organised as follows. Section~\ref{sec:background} introduces the background, and Section~\ref{sec:related-work} reviews the related work. Section~\ref{sec:motivating-scenarios} presents the motivating scenarios for our studies. Sections~\ref{sec:study-1} and~\ref{sec:study-2} describe the two empirical studies and present their results. Section~\ref{sec:cross-study-discussion} provides a cross-study discussion, and Section~\ref{sec:conclusion} concludes the paper.

\section{Background}\label{sec:background}
We now introduce the Large Language Models and prompt engineering strategies used in our study.

\subsection{Large Language Models}

\textit{Large Language Models (LLMs)} are neural networks trained on large text corpora~\cite{Hou2024,Fan2023}. They have demonstrated strong capabilities in reasoning, classification~\cite{binkhonain2025prompts}, summarization~\cite{DBLP:conf/refsq/MallyaFZD26}, and text generation~\cite{Arora2024}. These capabilities make them increasingly attractive for RE~\cite{zadenoori2025largelanguagemodelsllms}. Unlike traditional Natural Language Processing (NLP) and Machine Learning (ML) approaches, LLMs can be adapted to diverse RE tasks using natural-language prompts rather than task-specific models~\cite{Ferrari2025}.

LLMs can be broadly divided into two categories: \textit{lightweight LLMs} and \textit{frontier LLMs}~\cite{10.1145/3711896.3736563, zhao2026survey}. \textit{Lightweight LLMs} require fewer computational resources and can be executed locally~\cite{zadenoori2025largelanguagemodelsllms}. This makes them suitable for privacy-sensitive environments, controlled experimentation, and resource-constrained RE/SE projects~\cite{DBLP:journals/corr/abs-2510-21443}. However, their lower computational requirements come at the cost of reduced reasoning capabilities and smaller context windows compared with \textit{frontier LLMs}~\cite{10.1145/3711896.3736563}. In contrast, \textit{frontier LLMs} provide substantially stronger reasoning capabilities and larger context windows, enabling the analysis of long and heterogeneous software artefacts~\cite{zhao2026survey}. Their reliance on provider-hosted infrastructure, however, may introduce higher computational costs and data privacy considerations~\cite{DBLP:journals/corr/abs-2310-03533}. Consequently, the choice of model depends on the characteristics of the RE/SE task~\cite{zadenoori2025largelanguagemodelsllms}. \textit{Lightweight LLMs} are well suited to analysing relatively short artefacts such as user feedback, whereas \textit{frontier LLMs} are more appropriate for reasoning across multiple software artefacts, including project documentation, issue-tracking systems, and developer discussions.

To investigate LLM capabilities across different RE activities, this paper evaluates both lightweight and frontier models. \textbf{Study~I} investigates five lightweight open-source LLMs (Llama~2\footnote{\label{fn:llama2}\url{https://huggingface.co/meta-llama/Llama-2-7b-hf}}, Llama~3\footnote{\label{fn:llama3}\url{https://huggingface.co/meta-llama/Meta-Llama-3-8B}}, Mistral\footnote{\label{fn:mistral}\url{https://huggingface.co/mistralai/Mistral-7B-v0.3}}, Gemma~2\footnote{\label{fn:gemma2}\url{https://huggingface.co/google/gemma-2-9b}}, and Phi-3 Mini\footnote{\label{fn:phi3mini}\url{https://huggingface.co/microsoft/Phi-3-mini-4k-instruct}}) for feedback-driven RE tasks. \textbf{Study~II} evaluates two frontier LLMs (GPT-5.5\footnote{\label{fn:gpt55}\url{https://chatgpt.com/}}, and DeepSeek-R1\footnote{\label{fn:deepseek}\url{https://chat.deepseek.com/}}) for traceability analysis across software project artefacts.

The lightweight models were selected as a representative set of locally deployable LLMs from major AI developers. They differ in size and reasoning capability while remaining suitable for controlled local experimentation. Local execution provided full control over the experimental environment and ensured reproducibility. Rather than identifying a single best-performing model (e.g.,~\cite{DBLP:journals/corr/abs-2510-21443}), \textbf{Study~I} examines how representative lightweight LLMs perform across different RE tasks.

In contrast, \textbf{Study~II} evaluates frontier LLMs motivated by an industrial collaboration with Huawei Research Centre. Preliminary experimentation showed that lightweight models were impractical for traceability analysis over long and heterogeneous software artefacts because of their limited context windows and reasoning capabilities. Consequently, we evaluate GPT-5.5 and DeepSeek-R1, two representative frontier reasoning models with substantially larger context windows, accessed through their provider-hosted interfaces. These models represent complementary commercial and open-weight approaches to frontier LLM deployment in contemporary SE practice.

Table~\ref{tab:llms} summarises the LLMs evaluated in this paper. \textit{Parameters} denote the approximate number of trainable model weights (where publicly available), while the \textit{Context Window} indicates the maximum number of tokens the model can process in a single interaction. \textit{Category} distinguishes lightweight and frontier models, and \textit{Study} identifies the corresponding empirical study.

\begin{table}[t]
	\centering
	\caption{Large language models evaluated in this paper.}
	\label{tab:llms}
	\footnotesize
	\setlength{\tabcolsep}{5pt}
	\begin{tabular}{|l|l|c|c|c|c|}
		\hline
		\textbf{Model} & \textbf{Developer} & \textbf{Parameters} & \textbf{Context Window} & \textbf{Category} & \textbf{Study} \\
		\hline
		Llama 2 & Meta AI & 7B & 4K & Lightweight & I \\
		Llama 3 & Meta AI & 8B & 8K & Lightweight & I \\
		Mistral & Mistral AI & 7B & 8K & Lightweight & I \\
		Gemma 2 & Google DeepMind & 9B & 8K & Lightweight & I \\ 
		Phi-3 Mini & Microsoft & 3.8B & 4K & Lightweight & I \\ 
		\hline
		GPT-5.5 & OpenAI & Not disclosed & 128K & Frontier & II \\
		DeepSeek-R1 & DeepSeek AI & 671B & 128K & Frontier & II \\
		\hline
	\end{tabular}
\end{table}

\subsection{Prompting Strategies}
Interaction with LLMs is achieved through \textit{prompts}, i.e., natural-language instructions that define the task and expected output~\cite{binkhonain2025prompts,zadenoori2025largelanguagemodelsllms}. The formulation of a prompt, commonly referred to as a \textit{prompting strategy}, substantially influences model behaviour and output quality~\cite{DBLP:journals/corr/abs-2601-01954, zhao2026survey}. Consequently, prompt engineering has become an important aspect of applying LLMs to RE/SE  tasks~\cite{Huang2025PromptEF}. Appropriate prompting can improve reasoning quality, increase output consistency, and reduce ambiguity in model responses~\cite{vogelsang2024usinglargelanguagemodels}. These improvements are particularly important for RE tasks, where accurate interpretation and structured outputs are often required~\cite{zadenoori2025largelanguagemodelsllms}.

This paper considers five representative prompting strategies, summarised in Table~\ref{tab:prompt_strategies}. The strategies differ in how they guide model behaviour~\cite{binkhonain2025prompts, zhao2026survey,zadenoori2025largelanguagemodelsllms}. \textit{Zero-shot} prompting relies solely on task descriptions. \textit{Few-shot} prompting supplements the instruction with a small number of labelled examples. \textit{Chain-of-thought} prompting encourages explicit reasoning before producing the final answer. \textit{Constraint-based} prompting introduces explicit rules or templates to improve output consistency and structure. \textit{Role-based} prompting assigns the model a specific professional perspective to guide its responses.

We selected these prompting strategies because they are widely adopted in the LLM for RE/SE literature and cover the principal mechanisms for guiding model behaviour~\cite{zadenoori2025largelanguagemodelsllms, vogelsang2024usinglargelanguagemodels,DBLP:journals/corr/abs-2310-03533}. \textbf{Study~I} evaluates all five strategies to investigate their influence on classification and requirements generation tasks. In contrast, \textbf{Study~II} focuses on zero-shot and chain-of-thought prompting. Rather than optimising prompt design~\cite{DBLP:journals/corr/abs-2601-01954}, the study evaluates the reasoning capabilities of frontier LLMs over heterogeneous software project artefacts in a realistic industrial setting.

\begin{table}[h!]
	\centering
	\caption{Overview of prompt strategies used in our study.}
	\label{tab:prompt_strategies}
	\small
	\begin{tabular}{|p{3cm}|p{8.5cm}|}
		\hline
		\textbf{Prompt Strategy} & \textbf{Description and Example} \\
		\hline
		\multirow{4}{*}{\centering\textbf{Zero-shot}} &
		\textbf{Description:} A brief instruction describes the task, assuming the model can generalise from its pre-trained knowledge to perform it. \\
		\cline{2-2}
		& \textbf{Example:} ``Classify the following user feedback as a feature request, bug report, or usability issue.'' \\
		\hline		
		\multirow{4}{*}{\centering\textbf{Few-shot}} &
		\textbf{Description:} A few labelled examples show the desired pattern, assuming the model will apply it to new inputs. \\
		\cline{2-2}
		& \textbf{Example:} ``Example 1: `Add dark mode' $\rightarrow$ Feature Request. 
		Example 2: `App crashes on login' $\rightarrow$ Bug Report. 
		Now classify the next five items.'' \\
		\hline
		\multirow{4}{*}{\centering\textbf{Chain-of-Thought}} &
		\textbf{Description:} The model is instructed to reason step by step and write intermediate steps before giving the final answer. \\
		\cline{2-2}
		& \textbf{Example:} ``Explain why this discussion fragment supports the given project goal, then identify the traceability link.'' \\
		\hline
		\multirow{4}{*}{\centering\textbf{Constraint-based}} &
		\textbf{Description:} Prompts include explicit rules, templates, or conditions that the output must follow. \\
		\cline{2-2}
		& \textbf{Example:} ``Generate requirements that are testable, unambiguous, and measurable.'' \\
		\hline
		\multirow{4}{*}{\centering\textbf{Role-based}} &
		\textbf{Description:} The model is assigned a specific role and responds using the perspective, tone, and knowledge expected from it. \\
		\cline{2-2}
		& \textbf{Example:} ``You are a requirements analyst. Rewrite the following user request as a formal functional requirement.'' \\
		\hline
	\end{tabular}
	
\end{table}

\section{Related Works}\label{sec:related-work}
\JD{This section reviews related work on modern Requirements Engineering and LLMs for RE, highlighting the research gaps addressed in this study.

\subsection{Modern Requirements Engineering}

Requirements Engineering (RE) has evolved substantially over the past two decades~\cite{DBLP:journals/software/MaalejNJR16,8474507, DBLP:conf/enase/Franch20a}. Traditional RE relied primarily on stakeholder interviews, workshops, and requirements specifications~\cite{Nuseibeh2000}. Modern software projects, however, continuously generate large volumes of requirements-related information throughout the software lifecycle. This shift has led to the emergence of Data-Driven Requirements Engineering (DDRE)~\cite{DBLP:conf/icse/MaalejNR19, 10.1007/s42979-020-00416-4}. DDRE complements stakeholder knowledge with evidence extracted from heterogeneous software artefacts, such as user feedback~\cite{Dabrowski2022, DBLP:conf/re/MotgerOTFM25}, issue reports~\cite{DBLP:conf/re/WangHYZ24}, software repositories~\cite{DBLP:journals/is/Morales-Ramirez19}, and developer discussions~\cite{DBLP:journals/infsof/KifetewPSSMM21}.

The growing availability of software artefacts has stimulated research on Artificial Intelligence for Requirements Engineering (AI4RE)~\cite{10.1109/ASEW67777.2025.00053}. AI4RE applies Machine Learning (ML), information retrieval, knowledge-based techniques, and Natural Language Processing (NLP) to automate RE activities~\cite{Dabrowski2022,10.1007/s42979-020-00416-4}. These include, for example, requirements classification (e.g., functional versus non-functional requirements)~\cite{DBLP:conf/refsq/PatelMS25}, information extraction (e.g., feature descriptions from online user feedback)~\cite{DBLP:conf/refsq/ShahSSP25}, traceability recovery (e.g., linking requirements to source code)~\cite{DBLP:conf/re/PasslackSHEEGJS25}, and defect detection (e.g., identifying ambiguous requirements)~\cite{DBLP:journals/jss/AmnaWPHP25}.

As most software artefacts are represented in natural language, Natural Language Processing for Requirements Engineering (NLP4RE) has become a major research direction~\cite{DBLP:conf/icse/MaalejNR19}. NLP4RE extends traditional AI4RE by focusing on methods for understanding and processing textual software artefacts~\cite{Ferrari2025}. However, systematic literature reviews report that most AI- and NLP-based approaches address individual RE tasks, focus on a single artefact type (e.g., app reviews, issue reports, or requirements specifications)~\cite{10.1145/3444689,Sonbol2022TheUO}, and require task-specific models or labelled datasets for supervised learning~\cite{Dabrowski2022,zadenoori2025largelanguagemodelsllms,cheng2024generative}.

Recent advances in LLMs have significantly expanded these capabilities~\cite{cheng2024generative}. Unlike earlier AI and NLP approaches, LLMs support both analytical (e.g., requirements classification~\cite{DBLP:journals/corr/abs-2510-21443} and traceability recovery~\cite{10.1007/s00766-026-00460-1}) and generative RE activities (e.g., requirements specification~\cite{DBLP:conf/re/PasqualeRPD25} or model generation~\cite{DBLP:conf/refsq/ChouAD26}) through natural-language prompting~\cite{zadenoori2025largelanguagemodelsllms}. They can analyse diverse software artefacts and support multiple RE tasks using the same general-purpose model~\cite{https://doi.org/10.1002/spe.70029}.

Overall, the literature demonstrates a clear evolution of modern RE. Research has progressed from stakeholder-centric practices towards data-driven analysis of heterogeneous software artefacts~\cite{7888433,DBLP:conf/icse/MaalejNR19,10.1007/s42979-020-00416-4}. At the same time, AI and NLP approaches have evolved from task-specific techniques to general-purpose language models~\cite{Ferrari2025}. This evolution has laid the foundation for contemporary research on LLMs for RE~\cite{https://doi.org/10.1002/spe.70029, zadenoori2025largelanguagemodelsllms}.

\subsection{Large Language Models for Requirements Engineering}

Research on LLMs for RE has grown rapidly in recent years~\cite{zadenoori2025largelanguagemodelsllms,10.3389/fcomp.2025.1519437}. Existing studies investigate LLMs across a broad range of RE activities~\cite{cheng2024generative, Huang2025PromptEF}; and demonstrate their potential to support different stages of the requirements lifecycle. Recent systematic literature reviews also indicate a clear shift from traditional NLP pipelines towards prompt-driven and generative approaches (e.g., ~\cite{zadenoori2025largelanguagemodelsllms,cheng2024generative,Huang2025PromptEF,10.3389/fcomp.2025.1519437}). In particular, recent studies increasingly focus on requirements elicitation and validation~\cite{10.1007/978-3-031-88531-0_19}, while also exploring the integration of RE with downstream SE activities such as testing~\cite{11190385}, modelling~\cite{DBLP:conf/refsq/ChouAD26}, and design support~\cite{10.1007/978-3-031-88531-0_20}.

Current research covers both \textit{analytical} and \textit{generative} RE tasks~\cite{DBLP:conf/refsq/HessVFHKR26,zadenoori2025largelanguagemodelsllms}. Papers investigating \textit{analytical} tasks typically focus on requirements classification (e.g., functional versus non-functional requirements)~\cite{DBLP:conf/refsq/MallyaFZD26} and traceability recovery (e.g., linking related artefacts)~\cite{10.1007/s00766-026-00460-1}. In contrast, papers investigating \textit{generative} tasks commonly address requirements specification (e.g., generating user stories)~\cite{DBLP:conf/re/PasqualeRPD25}, refinement (e.g., improving requirement quality~\cite{10628462}, and explanation generation (e.g., explaining requirements)~\cite{10628462}.

Although the range of RE tasks has broadened considerably, most empirical studies still evaluate LLMs on a single task or application scenario~\cite{zadenoori2025largelanguagemodelsllms,cheng2024generative}. Recent work also increasingly explores more knowledge-intensive activities, including requirements elicitation~\cite{DBLP:conf/refsq/HessVFHKR26}, validation~\cite{10.1007/978-3-031-88531-0_19}, and specification generation~\cite{10.3389/fcomp.2025.1519437}. Consequently, it remains unclear to what extent findings obtained for one RE task generalise to others~\cite{zadenoori2025largelanguagemodelsllms,Huang2025PromptEF}. This limits a broader empirical understanding of LLM capabilities across RE tasks.

The range of evaluated artefacts has also expanded considerably. Early studies focused primarily on Software Requirements Specifications (SRS)~\cite{DBLP:conf/re/PasqualeRPD25,zadenoori2025largelanguagemodelsllms}. More recent work investigates a broader range of software artefacts, including user stories~\cite{DBLP:conf/refsq/MallyaFZD26}, issue reports~\cite{10.1007/s00766-026-00462-z}, online user feedback~\cite{11190331}, regulatory documents, technical documentation~\cite{Huang2025PromptEF}, and software repositories~\cite{zadenoori2025largelanguagemodelsllms}. This trend reflects the increasing importance of heterogeneous software artefacts in modern RE.

Empirical studies investigate a growing range of LLMs and prompting strategies~\cite{10.1007/s00766-026-00460-1,10.1007/s00766-026-00462-z}. However, most evaluations still rely on frontier models (e.g., GPT-based) and zero-shot or few-shot prompting~\cite{zadenoori2025largelanguagemodelsllms}. Lightweight models (e.g., Mistral) and more advanced prompting strategies, including chain-of-thought reasoning~\cite{DBLP:journals/corr/abs-2510-21443}, retrieval-augmented generation (RAG)~\cite{10.1007/978-3-031-88531-0_20}, and agent-based workflows~\cite{11190353}, have received comparatively less attention. Studies are also typically conducted either as controlled experiments~\cite{DBLP:journals/corr/abs-2510-21443} or as industrial case studies~\cite{DBLP:conf/re/PasqualeRPD25}, with few combining both evaluation settings.

\subsection{Research Gap}
\begin{table}[t]
	\centering
	\caption{Comparison between existing empirical studies on LLMs for Requirements Engineering and this work.}
	\label{tab:relatedwork}
	\begin{tabular}{|p{2.5cm}|p{4.9cm}|p{4.4cm}|}
		\hline
		\textbf{Dimension} & \textbf{Existing Studies} & \textbf{This Work} \\
		\hline
		Evaluation scope &
		Single RE task &
		Multiple RE tasks \\
		\hline
		RE task type &
		Analytical \textit{or} generative &
		Analytical \textit{and} generative \\
		\hline
		Evaluation setting &
		Experiment \textit{or} case study &
		Experiment \textit{and} case study \\
		\hline
		Models &
		Single LLM family &
		Lightweight and frontier LLMs \\
		\hline
		Reproducibility &
		Limited replication support &
		Replication package \\
		\hline
	\end{tabular}
\end{table}

Recent AI4RE studies demonstrate the growing potential of LLMs for RE applications~\cite{zadenoori2025largelanguagemodelsllms,cheng2024generative,Huang2025PromptEF,10.3389/fcomp.2025.1519437}. However, the available empirical evidence remains fragmented~\cite{Huang2025PromptEF,10.3389/fcomp.2025.1519437}. Table~\ref{tab:relatedwork} summarises the principal differences between previous empirical studies and our work.

First, most studies investigate a single RE task (e.g., requirements classification~\cite{DBLP:conf/refsq/PatelMS25} or traceability recovery~\cite{10.1007/s00766-026-00460-1}), limiting the generalisability of existing findings across different RE tasks.

Second, previous studies typically evaluate either analytical or generative RE tasks, and rely on either controlled experiments or industrial case studies~\cite{cheng2024generative,zadenoori2025largelanguagemodelsllms}. As a result, technical performance and practical applicability are rarely assessed together. In addition, most evaluations consider models from a single LLM family~\cite{zadenoori2025largelanguagemodelsllms}, providing limited insight into the trade-offs between lightweight and frontier LLMs~\cite{DBLP:journals/corr/abs-2510-21443}.

Finally, reproducibility remains an important challenge~\cite{zadenoori2025largelanguagemodelsllms,vogelsang2024usinglargelanguagemodels}. While some studies release datasets, prompts, source code, or other research artefacts, replication support is often incomplete or unavailable~\cite{DBLP:journals/tosem/AbualhaijaADDFFF24}. This limits independent validation, cross-study comparisons, and the accumulation of reliable empirical evidence.

To address these gaps, our study provides a comprehensive cross-task evaluation of LLMs for RE. It evaluates multiple analytical and generative RE tasks using both lightweight and frontier LLMs, combines a controlled experiment with an industry-motivated case study, and releases a complete replication package to support reproducibility and future research.
}

\section{Motivating Scenarios}\label{sec:motivating-scenarios}
We now present two complementary RE scenarios representing different stages of the requirements lifecycle. The first scenario is motivated by prior empirical RE research on the use of online user feedback~\cite{Dabrowski2022}. Specifically, it builds on empirical studies, surveys, and practitioner interviews examining how software teams leverage user feedback to support RE and SE practice~\cite{AlSubaihin2021}. The second scenario is motivated by an industry-informed case study conducted in collaboration with Huawei Research Centre, focusing on the Rust open-source ecosystem\footnote{\url{https://rust-lang.org}}. It investigates requirements-related information distributed across heterogeneous software project artefacts to support requirements traceability, design rationale understanding, and impact analysis~\cite{Ferrari2025}. Table~\ref{tab:motivating_scenarios} provides an overview of the motivating scenarios, the supported RE activities, and the corresponding empirical studies.

\begin{table}[t]
	\centering
	\caption{Motivating scenarios and supported RE activities.}
	\label{tab:motivating_scenarios}
	\small
	\begin{tabular}{|p{5.5cm}|p{5.2cm}|c|}
		\hline
		\textbf{Motivating Scenario} &
		\textbf{Supported RE Activities} &
		\textbf{Study} \\
		\hline
		From User Feedback to Requirements &
		\begin{tabular}[c]{@{}l@{}}
			Requirements elicitation\\
			Requirements classification\\
			Requirements specification generation
		\end{tabular}
		& I \\
		\hline
		Requirements Across Project Artefacts &
		\begin{tabular}[c]{@{}l@{}}
			Requirements traceability\\
			Design rationale understanding\\
			Requirements impact analysis
		\end{tabular}
		& II \\
		\hline
	\end{tabular}
\end{table}

\subsection{Scenario 1: From User Feedback to Requirements}

Modern mobile applications receive large volumes of online user reviews through app store platforms~\cite{Dabrowski2022}. These reviews contain valuable requirements-related information, including feature requests, bug reports, and software quality concerns (e.g., usability and performance)~\cite{DBLP:journals/software/MaalejNJR16}. Analysing user reviews can supports several RE activities~\cite{Dabrowski2022a}. However, their volume and noisy nature make manual analysis challenging~\cite{Dabrowski2022}. Consequently, automated approaches for analysing user feedback and generating requirements specifications are increasingly important for supporting RE practice~\cite{DBLP:conf/icse/MaalejNR19}.

Consider a development team maintaining a mobile application such as WhatsApp. Following a new release, the team receives thousands of user reviews. Some request new functionality (e.g., ``add message scheduling''), others report defects (e.g., ``messages fail to send''), while many describe software quality concerns such as poor performance (``the application is too slow to open'') or usability issues (``the chat layout is confusing''). Although valuable, only a subset of reviews contains actionable requirements-related information.

To identify stakeholder needs, the team could automatically classify user reviews by request type (e.g., feature request, bug report, or other) and referenced non-functional requirements (NFRs). The resulting classifications would support \emph{requirements elicitation} by extracting actionable stakeholder needs from large volumes of user feedback. They would support \emph{requirements classification} by organising the elicited requirements according to request type and quality attributes. By quantifying the classified feedback, developers could identify the most frequently requested features, reported defects, and software quality concerns. This would support \emph{requirements prioritisation} and guide software evolution~\cite{Dabrowski2022,Dabrowski2022a}.

Having identified and classified the relevant feedback, the team could support the \emph{requirements specification} activity. An automated approach could generate requirement statements, user stories, or an initial Software Requirements Specification (SRS). For example, the review ``add dark mode'' could be reformulated as the requirement statement ``The system shall provide a dark mode option.'' Alternatively, it could be expressed as the user story ``As a user, I want to enable dark mode so that I can comfortably use the application in low-light environments.'' Although these drafts would not replace a complete specification, they could reduce documentation effort, improve consistency, and maintain traceability between user feedback and documented requirements~\cite{AlSubaihin2021}.

This scenario motivates \textbf{Study I}, which investigates lightweight open-source LLMs for supporting these RE activities.

\subsection{Scenario 2: Requirements Across Project Artefacts}

As software projects evolve, requirements-related information becomes distributed across software repository artefacts, including project roadmaps, GitHub issues, and developer discussions~\cite{DBLP:conf/enase/Franch20a,DBLP:journals/software/MaalejNJR16}. Analysing the relationships between these artefacts supports RE activities such as requirements traceability, design rationale understanding, and requirements impact analysis~\cite{Ferrari2025}. However, their distributed and heterogeneous nature makes manual analysis challenging~\cite{10.1007/s42979-020-00416-4}. Consequently, automated traceability identification and rationale explanation are increasingly important for RE practice~\cite{10.1007/s00766-026-00460-1}.

Consider the real-world open-source Rust programming language project, where contributors from both the community and industry (e.g., Huawei) collaborate on the language's development\footnote{\label{fn:project_rust}\url{https://github.com/rust-lang/}}. Requirements-related information is distributed across roadmap goals\footnote{\label{fn:project_rust_goal}\url{https://rust-lang.github.io/rust-project-goals/}}, GitHub issues\footnote{\label{fn:project_rust_issue}\url{https://github.com/rust-lang/rust/issues/}}, and Zulip developer discussions\footnote{\label{fn:project_rust_zulip}\url{https://rust-lang.org/community/}}. Understanding the relationships between these artefacts would help contributors answer questions such as: \emph{Which roadmap goal does this GitHub issue address? What functionality is being proposed? Why was this design decision made? How should the proposed feature be implemented?} Answering these questions supports requirements traceability, design rationale understanding, and anticipating the impact of proposed language changes~\cite{Ferrari2025}. However, manually reconstructing these relationships across numerous artefacts is time-consuming and error-prone.

The team could therfore use an automated approach to identify traceability links and generate evidence-grounded explanations across project artefacts. The identified links would support \emph{requirements traceability} by linking roadmap goals with related GitHub issues and Zulip discussions. Contributors could then quickly determine which discussions concern a particular project goal, which issue operationalises it, and how the proposed feature evolves throughout development. The generated explanations would support \emph{design rationale understanding} by explaining why these artefacts are linked and identifying the evidence supporting each relationship.

When planning language changes, contributors need to anticipate their implications before implementation. Traceability links and their evidence-grounded explanations would help identify related roadmap goals, implementation activities, and technical dependencies that may be affected by a proposed feature or modification. The same information would also help industrial adopters assess how language evolution could influence downstream software systems, such as operating systems, embedded software, and Rust-based applications. This would support \emph{requirements impact analysis} and enable better-informed engineering decisions.

This scenario motivates \textbf{Study II}, which investigates state-of-the-art LLMs for supporting these RE activities across heterogeneous software project artefacts.

These two scenarios cover complementary stages of the RE lifecycle. Scenario I focuses on eliciting and documenting requirements from stakeholder feedback, whereas Scenario II focuses on tracing and explaining requirements across software project artefacts.

\section{Study I: From User Feedback to Requirements}\label{sec:study-1}
The first study investigates whether lightweight LLMs can support feedback-driven RE, based on the first motivating scenario (see Sect.~\ref{sec:motivating-scenarios}). It evaluates five lightweight open-source LLMs across three complementary tasks: user request classification, NFR classification, and requirements specification generation. We first introduce the study context, including the key concepts, terminology, and RE problems considered in this study. We then present the empirical study design, experimental findings, and discuss their implications for RE practice.

\subsection{Study Context}
Our first study focuses on \textit{app review analysis}~\cite{Dabrowski2022}. Specifically, we consider \textit{app reviews}, a form of online user feedback collected from mobile application platforms (e.g., Google Play Store). App reviews are among the most widely used sources of user feedback in RE. Their large volume and unstructured nature make manual analysis challenging, motivating automated approaches for user feedback analysis~\cite{AlSubaihin2021,11190331}. Each app review is treated as a \textit{user feedback instance}.

User feedback conveys diverse types of information relevant to RE. We focus on two information types: \textit{user requests} and \textit{non-functional requirements (NFRs)}. User requests express users' needs to modify the software and are categorized as either \textit{feature requests}, describing desired functionality, or \textit{bug reports}, describing defects or incorrect system behavior~\cite{Dabrowski2022}. NFRs are categorized according to the software quality attributes defined by the ISO/IEC~25010 quality model, including usability, reliability, performance, and security~\cite{iso25010}.

We investigate two complementary RE tasks. The first, \textit{user feedback classification}, classifies each app review using two schemes: the \textit{user request type} (e.g., \textit{feature request} or \textit{bug report}) and the \textit{NFR type} (e.g., \textit{performance}, \textit{usability}, or \textit{security}), each with an additional \textit{other} category. The second, \textit{requirements specification generation}, transforms one or more classified user feedback instances into structured requirements specifications, such as requirement statements or user stories, that capture the underlying stakeholder needs.

For example, the review \textit{`The app crashes when saving a form''} is classified as a \textit{bug report}, whereas \textit{`I want the app to automatically save my progress''} is classified as a \textit{feature request}. Similarly, \textit{`The app responds very slowly when opening large files''} is classified into the \textit{performance} NFR category. The feature request can then be transformed into the requirement statement \textit{`The system shall automatically save user progress.''} or expressed as the user story \textit{``As a user, I want the app to automatically save my progress so that I do not lose data.''}


\subsection{Empirical Study Design}
\label{sec:design-1}
This section presents the empirical study conducted to evaluate the effectiveness of LLMs in analysing online user feedback to support RE. 

\subsubsection{Research Questions}
\label{sec:RQs}
The goal of this study is to evaluate LLMs in analysing online user feedback to support RE tasks. We specifically focus on three research questions:

\begin{itemize}
	\item \textbf{RQ1}: How well do LLMs classify feedback by \JAC{NFR type?}
	\item \textbf{RQ2}: How well do LLMs classify feedback by user-request type?
	\item \textbf{RQ3}: How well do LLMs generate requirements specification?
\end{itemize}

In RQ1, we assess the models’ ability to identify the NFR type mentioned in user reviews. RQ2 examines how accurately the models classify user feedback by request type. RQ3 evaluates their capability to generate requirements specification from the same feedback. All evaluations use human-annotated datasets (see Sect.~\ref{sec:datasets}). For RQ1–RQ2, model predictions are compared with annotations using precision, recall, and F1-score. For RQ3, SRSs generated from sampled reviews are assessed through human judgment based on predefined quality criteria.

\subsubsection{Datasets}
\label{sec:datasets}

\JAC{We use two annotated datasets of mobile app user feedback from prior studies \cite{jha2018using,lu2017automatic}. The first dataset was originally labeled with NFR types \cite{lu2017automatic} and is used to answer RQ1. The second dataset contains app reviews annotated with user request types \cite{jha2018using} and is used to answer RQ2. In addition, a sample of annotated reviews from both datasets is used to answer RQ3. We selected these datasets for their relevance and availability in a public repository \cite{Dabrowski2022}. Each dataset contains thousands of reviews from approximately a dozen apps across diverse domains and both major app stores. This diversity mitigates the app sampling problem and supports validity \cite{Dabrowski2022}.}

\noindent\textbf{Non-Functional Requirements (NFR) Dataset.} The first dataset, introduced by Lu and Liang~\cite{lu2017automatic}, focuses on app reviews annotated with NFR types based on the software quality model~\cite{iso25010}. The initial collected dataset comprised approximately 11,000 app reviews from two mobile applications in the books and communications categories. These reviews were drawn from both major app stores, the Apple App Store and Google Play Store. A subset of 4,000 review sentences was manually annotated according to five NFR categories.

\noindent\textbf{User Request Dataset.} The second dataset builds on previous studies~\cite{chen2014arminer,maalej2016on} and includes additional user feedback collected for this study~\cite{jha2018using}. The initial collected dataset covered about 10 mobile applications across more than 10 categories from both the Apple App Store and Google Play Store. From an initial pool of the collected reviews, a curated subset of 2,912 was manually annotated into three user request types: feature requests, bug reports, and other. This dataset provides a diverse and representative sample of user feedback for evaluating automated classification approaches.

\subsubsection{Evaluation Metrics and Criteria}
\label{sec:metrics}

We applied both quantitative and qualitative methods. Standard ML metrics~\cite{11190353} were used for RQ1 and RQ2, as feedback classification is a classification task~\cite{geron2022hands}. Specification generation (RQ3) was evaluated via criteria-based assessment~\cite{kuckartz2014qualitative}.

\paragraph{Evaluation Metrics for Classification (RQ1--RQ2).} We compute precision, recall, and F1-score for two experiments: classifying user feedback by NFR type (RQ1) and by user request type (RQ2). Precision measures the proportion of correctly predicted labels among all predictions, while recall measures the proportion of correctly identified labels in the ground truth. Model performance is assessed by comparing each predicted label (user request or NFR type) with its annotated counterpart. Scores are calculated per class (e.g., feature request, bug report, other), along with macro averages. The macro average treats all classes equally.

\paragraph{Evaluation Criteria for Specification Generation (RQ3).} For RQ3, the quality of generated requirement specification is evaluated qualitatively across \JAC{five} criteria derived from established specification attributes such as completeness, consistency, and clarity~\cite{10834143}. Each criterion was adapted to suit the characteristics of automatically generated specification. Table~\ref{tab:srs_rubric} outlines the evaluation rubric, which includes \JAC{five} criteria. These dimensions assess structural correctness, coverage of stakeholder input, factual grounding, semantic accuracy, and linguistic quality. Each criterion follows a defined scoring scheme combining quantitative (e.g., counts or coverage) and qualitative (e.g., 1–5 scale) measures, supporting systematic and replicable assessment of specification quality~\cite{kuckartz2014qualitative}.

\begin{table}[t]
	\centering
	\caption{\JD{Evaluation criteria and scoring used to assess generated specification (RQ3).}}
	\label{tab:srs_rubric}
	\small
		\begin{tabular}{|p{3.6cm}|p{9cm}|}
			\hline
			\textbf{Criterion} & \textbf{Description and Scoring} \\
			\hline
			
			\multirow{4}{*}{\centering\textbf{Structural Adherence}} &
			\textbf{Description:} Evaluates how well the generated specification follows the expected structure, ensuring coverage of all sections. \\
			\cline{2-2}
			& \textbf{Scoring:} \JD{Originally 1–8 points (one per correctly included section), linearly rescaled to 1–5 for consistency; higher is better.} \\
			\hline
			
			\multirow{4}{*}{\centering\textbf{Completeness}} &
			\textbf{Description:} Evaluates the coverage of requirements identified from user feedback. \\
			\cline{2-2}
			& \textbf{Scoring:} Rated on a 1–5 scale, with higher scores reflecting greater completeness of identified requirements from user feedback. \\
			\hline
			
			\multirow{4}{*}{\centering\textbf{Fidelity}} &
			\textbf{Description:} \JD{Evaluates how faithfully the generated specification reflects user feedback, identifying fabricated requirements.} \\
			\cline{2-2}
			&  \JD{\textbf{Scoring:} Rated on a 1–5 scale; higher scores indicate greater fidelity (rescaled from the proportion of fabricated to valid requirements).}	\\
			\cline{2-2}
			\hline
			\multirow{4}{*}{\centering\textbf{Conciseness}} &
			\textbf{Description:} Evaluates redundancy and verbosity in the generated text, indicating how efficiently information is conveyed. \\
			\cline{2-2}
			& \textbf{Scoring:} \JD{Rated on a 1–5 scale, with higher scores reflecting greater conciseness and lower values indicate increased verbosity.}\\
			\hline
			\multirow{4}{*}{\centering\textbf{Clarity}} &
			\textbf{Description:} Evaluates the clarity of generated requirements in terms of unambiguity, specificity, and interpretability. \\
			\cline{2-2}
			& \textbf{Scoring:} Rated on a 1–5 scale, with higher scores reflecting greater clarity and linguistic precision. \\
			\hline
			
		\end{tabular}
	
\end{table}

\subsubsection{Experimental Setup and Procedure}
\label{sec:setup}

We now describe the computational setup, prompting strategies, and evaluation procedures used in three experiments (RQ1--RQ3).

\paragraph{Computational Setup}
\label{sec:comp-setup}
All experiments were conducted under consistent hardware and parameter settings to ensure fairness and reproducibility. We evaluated five LLMs (see Sect.~\ref{sec:background}), running each model three times per task to reduce stochastic variance. Default parameters were used, with \textit{temperature} set to 0 and a fixed \textit{random seed} to minimise non-determinism. No hyperparameter tuning was applied to isolate the effects of prompting strategies. Experiments ran on a workstation with an NVIDIA RTX 4050 GPU (6 GB VRAM) and 16 GB RAM. Total runtime per model (3 runs) ranged from 20 min to 2 h for user request classification (512 reviews) and 2–5 hours for NFR classification (1,278 reviews).

\paragraph{Prompting Strategies}
\label{sec:prompts} 

We experiment with five prompting strategies (see Sect.~\ref{sec:background}). Prompting strategies are customised for each experiment (RQ1--RQ2).\newline

\noindent\textit{Prompting Strategies for  Classification (RQ1--RQ2).} We apply three prompting strategies across two classification tasks (user request type and NFR type): \textit{zero-shot}, \textit{few-shot}, and \textit{chain-of-thought (CoT)} prompting. These strategies capture increasing levels of reasoning and contextualization while remaining lightweight and reproducible. We omit more complex prompting (e.g., role-based or constraint-based), as classification primarily requires consistent label prediction rather than creative or constrained generation. Prompts are refined iteratively through a pilot study. Few-shot examples come from our dataset, and CoT prompts direct models to ``\textit{think step-by-step before giving the final category}''.\newline

\noindent\textit{Prompting Strategies for  Specification Generation (RQ3).} We use all five prompting strategies for the specification generation task. Beyond the three classification strategies (\textit{zero-shot}, \textit{few-shot}, \textit{CoT}), we add \textit{constraint-based} and \textit{role-based} prompting to improve structural coherence and contextual relevance. These strategies better suit generative tasks that require creativity and controlled output. In the constraint-based setup, prompts specify that outputs follow a defined structure (e.g., functional and non-functional sections) and include constraints such as ``\textit{avoid implementation details}'' and ``\textit{ensure each requirement is unique}''.

\paragraph{Evaluation Procedures}
\label{sec:eval-proc}

We use quantitative evaluation for classification tasks (RQ1--RQ2) and qualitative evaluation for specification generation (RQ3).\newline

\noindent\textit{Evaluation Procedure for Classification (RQ1--RQ2).} For RQ1 and RQ2, we evaluate models on the corresponding annotated dataset for each classification experiment (Section~\ref{sec:datasets}). Each review is processed by the LLM under each prompting setup (zero-shot, few-shot, CoT), and predicted labels are compared with the ground truth. We calculate precision, recall, and F1-score per class, along with macro- and weighted averages. We report mean values over three runs.\newline

\noindent\textit{Evaluation Procedure for Specification Generation (RQ3).} For RQ3, we use 90 annotated reviews \JAC{randomly} sampled from our collected data. Half are taken from the user request dataset, and the other half from the NFR dataset. The same input is provided to each LLM under every prompting setup. Each model generates requirements specification following the template, including Introduction, Functional Requirements, NFRs, and Glossary sections. \JD{The first author evaluates the outputs using the five criteria (see Table~\ref{tab:srs_rubric}). The results are verified through manual examination of user feedback. The scores are averaged across all samples.}

\subsection{Results}
\label{sec:results-1}
\JD{\noindent\textbf{RQ1: How well do LLMs classify feedback by NFR type?}

\noindent To answer RQ1, we evaluated how well LLMs classify user feedback by NFRs under three prompting strategies: zero-shot, few-shot, and chain-of-thought. Table \ref{tab:llm-nfrs-performance} reports precision, recall, and F1 scores for each model, with the best results highlighted in bold. The effectiveness ranges from an F1 score of 0.40 to 0.55, with averages of 0.47, 0.49, and 0.51 for the zero-shot, few-shot, and chain-of-thought strategies, respectively.  Across prompting strategies, performance improves steadily from zero-shot to few-shot to chain-of-thought, confirming that example-based and reasoning-enhanced prompts help LLMs better identify NFR types. Larger and newer models (Gemma, Llama 3, Mistral) generally outperform smaller or earlier ones (Llama 2, Phi-3 Mini). Gemma achieves the highest overall F1 score (0.55) and precision (0.53) under the chain-of-thought setting, while Llama 3 records the highest recall (0.59). Mistral performs most consistently across all prompting strategies, ranking near the top in zero-shot and few-shot modes. In contrast, Llama 2 yields the lowest scores across all metrics, while Phi-3 Mini remains stable but below the larger models.

\rqanswerfirst{LLMs achieve moderate accuracy (F1 $\approx$  0.47–0.51) for classifying feedback by NFR type. Chain-of-thought prompting performs best, with Gemma leading overall (P=0.53, R=0.57, F1=0.55).}

\begin{table*}[h]
	\centering
	\caption{How well do LLMs classify feedback by NFR type? (RQ1)}
	\label{tab:llm-nfrs-performance}
	\setlength{\tabcolsep}{6pt} 
	\renewcommand{\arraystretch}{1.2} 
	\scalebox{0.9}{
		\begin{tabular}{|l|c|c|c|c|c|c|c|c|c|}
			\hline
			\multirow{2}{*}{\textbf{Model}} 
			& \multicolumn{3}{c|}{\textbf{Zero-Shot}} 
			& \multicolumn{3}{c|}{\textbf{Few-Shot}} 
			& \multicolumn{3}{c|}{\textbf{Chain-of-Thought}} \\
			\cline{2-10}
			& \textbf{P} & \textbf{R} & \textbf{F1}
			& \textbf{P} & \textbf{R} & \textbf{F1}
			& \textbf{P} & \textbf{R} & \textbf{F1} \\
			\hline
			Llama 2        & 0.44 & 0.36 & 0.40 & 0.49 & 0.39 & 0.43 & 0.52 & 0.46 & 0.49 \\
			Llama 3        & 0.42 & \textbf{0.54} & 0.47 & 0.46 & \textbf{0.57} & 0.51 & 0.48 & \textbf{0.59} & 0.53 \\
			Mistral        & 0.49 & 0.51 & \textbf{0.50} & \textbf{0.54} & 0.51 & \textbf{0.52} & 0.47 & \textbf{0.59} & 0.52 \\
			Gemma          & \textbf{0.51} & 0.48 & 0.49 & 0.51 & 0.49 & 0.50 & \textbf{0.53} & 0.57 & \textbf{0.55} \\
			Phi-3 Mini     & 0.44 & 0.54 & 0.48 & 0.46 & 0.54 & 0.50 & 0.44 & 0.48 & 0.46 \\
			\hline
			\textbf{Average} & 0.46 & 0.49 & 0.47 
			& 0.49 & 0.50 & 0.49 
			& 0.49 & 0.54 & 0.51 \\
			\hline
		\end{tabular}
	}
\end{table*}

\noindent \textbf{RQ2: How well do LLMs classify feedback by user request type?}

\noindent To answer RQ2, we examined how LLMs classify user feedback by request type under three prompting strategies: zero-shot, few-shot, and chain-of-thought. Table \ref{tab:llm-ur-performance} presents precision, recall, and F1 scores for each model, with the best results in bold. Model performance ranges from an F1 score of 0.32 to 0.74. The average F1 values are 0.59 for zero-shot, 0.68 for few-shot, and 0.64 for chain-of-thought prompting. Performance increases notably from zero-shot to few-shot prompting, while chain-of-thought yields moderate improvements. Larger and more recent models, such as Llama 3, Mistral, and Gemma, generally achieve higher accuracy than smaller or earlier ones, including Llama 2 and Phi-3 Mini. Llama 3 reaches the highest F1 score of 0.74 and recall of 0.75 under the few-shot setting. Mistral and Gemma perform consistently well across all strategies. Llama 2 produces the lowest scores, while Phi-3 Mini remains stable but below the stronger models. Overall, few-shot prompting provides the best results (average F1 = 0.68); it suggests that including example-based context helps LLMs more accurately classify user feedback by request type.

\rqanswersec{LLMs achieve moderate-to-high average accuracy (F1 $\approx$ 0.59–0.68) for classifying feedback by user request type. Few-shot prompting performs best, with Llama 3 leading overall (P=0.72, R=0.75, F1=0.74).}

\begin{table*}[h]
	\centering
	\caption{How well do LLMs classify user feedback by request type? (RQ2)}
	\label{tab:llm-ur-performance}
	\setlength{\tabcolsep}{6pt} 
	\renewcommand{\arraystretch}{1.2} 
	\scalebox{0.9}{
		\begin{tabular}{|l|c|c|c|c|c|c|c|c|c|}
			\hline
			\multirow{2}{*}{\textbf{Model}} 
			& \multicolumn{3}{c|}{\textbf{Zero-Shot}} 
			& \multicolumn{3}{c|}{\textbf{Few-Shot}} 
			& \multicolumn{3}{c|}{\textbf{Chain-of-Thought}} \\
			\cline{2-10}
			& \textbf{P} & \textbf{R} & \textbf{F1}
			& \textbf{P} & \textbf{R} & \textbf{F1}
			& \textbf{P} & \textbf{R} & \textbf{F1} \\
			\hline
			Llama 2        & 0.28 & 0.36 & 0.32 & 0.57 & 0.56 & 0.57 & 0.60 & 0.42 & 0.49 \\
			Llama 3        & \textbf{0.77} & 0.67 & \textbf{0.72} & 0.72 & 0.75 & \textbf{0.74} & 0.71 & 0.70 & 0.71 \\
			Mistral        & 0.60 & 0.63 & 0.62 & 0.69 & \textbf{0.74} & 0.71 & 0.65 & \textbf{0.72} & 0.68 \\
			Gemma          & 0.66 & \textbf{0.71} & 0.68 & 0.68 & 0.67 & 0.68 & 0.65 & 0.60 & 0.63 \\
			Phi-3 Mini     & 0.69 & 0.51 & 0.59 & 0.68 & 0.67 & 0.68 & 0.67 & 0.70 & \textbf{0.69} \\
			\hline
			\textbf{Average} & 0.60 & 0.58 & 0.59 
			& 0.67 & 0.68 & 0.68 
			& 0.66 & 0.63 & 0.64 \\
			\hline
		\end{tabular}
	}
\end{table*}

\noindent \textbf{RQ3: How well do LLMs generate requirements specification?}

\noindent To answer RQ3, each model was evaluated on five criteria: structure (SA), completeness (CO), fidelity (FI), conciseness (CN), and clarity (CL), as defined in Table~\ref{tab:srs_rubric}. Table~\ref{tab:llm-srs-performance} reports results across models and prompting strategies. Overall, the models produced moderate-quality specification (mean=3.1; SD=0.8). Llama 3 with chain-of-thought prompting and Mistral with few-shot prompting achieved the highest mean score (3.6), showing strong structure and clarity (SA=4; CL=5). Llama 2 followed with stable mid-range scores (mean=3.2–3.4). Gemma produced the weakest but most consistent outputs (mean $\approx$ 2.9; SD=0.4). Phi-3 Mini showed high fidelity and conciseness (FI $\approx$ 4; CN=4–5) but low structure (SA=1–2) and high variability. Across prompting strategies, few-shot and chain-of-thought improved structure and clarity, whereas constraint-based prompting offered limited gains. Most models scored highest in clarity (CL=4–5) and completeness (CO=4) but lagged with fidelity and conciseness (FI, CN=2–3).

\rqanswerthree{LLMs produced moderate-quality specification, with Llama 3 and Mistral performing best; model and prompt choice strongly influenced output quality.}

\begin{table*}[h]
	\centering
\caption{LLM performance on requirement specification (RQ3). 
	SA – Structure; CO – Compl.; FI – Fidelity; CN – Conciseness; CL – Clarity. 
	Scores are on a 1–5 scale (higher = better). 
	\JAC{\textbf{Bold} indicates the best overall performance across all models and prompt strategies, based on the highest Mean score (averaged over the five criteria) and the lowest SD.}}

	\label{tab:llm-srs-performance}
	\setlength{\tabcolsep}{4pt}
	\renewcommand{\arraystretch}{1.2}
	\scalebox{0.76}{
		\begin{tabular}{|l|l|c|c|c|c|c|c|c|}
			\hline
			\textbf{Model} & \textbf{Prompt Strategy} & 
			\textbf{SA (1–5)} & 
			\textbf{CO (1–5)} & 
			\textbf{FI (1–5)} & 
			\textbf{CN (1–5)} & 
			\textbf{CL (1–5)} & 
			\textbf{Mean} & 
			\textbf{SD} \\ 
			\hline
			
			\multirow{5}{*}{Llama 2} 
			& Zero-Shot & 3 & 4 & 3 & 3 & 4 & 3.4 & 0.49 \\ \cline{2-9}
			& Few-Shot & 3 & 4 & 3 & 2 & 4 & 3.2 & 0.75 \\ \cline{2-9}
			& Chain-of-Thought & 3 & 4 & 3 & 2 & 4 & 3.2 & 0.75 \\ \cline{2-9}
			& Constraint-based & 3 & 4 & 3 & 2 & 4 & 3.2 & 0.75 \\ \cline{2-9}
			& Role-based & 3 & 4 & 3 & 2 & 4 & 3.2 & 0.75 \\ 
			\hline
			
			\multirow{5}{*}{Llama 3} 
			& Zero-Shot & 3 & 4 & 3 & 2 & 4 & 3.2 & 0.75 \\ \cline{2-9}
			& Few-Shot & 3 & 4 & 3 & 2 & 4 & 3.2 & 0.75 \\ \cline{2-9}
			& Chain-of-Thought & 4 & 4 & 3 & 2 & 5 & \textbf{3.6} & 1.02 \\ \cline{2-9}
			& Constraint-based & 3 & 4 & 3 & 2 & 4 & 3.2 & 0.75 \\ \cline{2-9}
			& Role-based & 3 & 4 & 3 & 2 & 4 & 3.2 & 0.75 \\ 
			\hline
			
			\multirow{5}{*}{Mistral} 
			& Zero-Shot & 3 & 3 & 3 & 2 & 4 & 3.0 & 0.63 \\ \cline{2-9}
			& Few-Shot & 4 & 4 & 3 & 2 & 5 & \textbf{3.6} & 1.02 \\ \cline{2-9}
			& Chain-of-Thought & 3 & 4 & 3 & 2 & 4 & 3.2 & 0.75 \\ \cline{2-9}
			& Constraint-based & 3 & 4 & 3 & 2 & 4 & 3.2 & 0.75 \\ \cline{2-9}
			& Role-based & 3 & 4 & 3 & 2 & 4 & 3.2 & 0.75 \\ 
			\hline
			
			\multirow{5}{*}{Gemma} 
			& Zero-Shot & 2 & 3 & 3 & 3 & 3 & 2.8 & \textbf{0.40} \\ \cline{2-9}
			& Few-Shot & 3 & 3 & 3 & 2 & 4 & 3.0 & 0.63 \\ \cline{2-9}
			& Chain-of-Thought & 3 & 3 & 3 & 2 & 4 & 3.0 & 0.63 \\ \cline{2-9}
			& Constraint-based & 2 & 3 & 3 & 2 & 4 & 2.8 & 0.75 \\ \cline{2-9}
			& Role-based & 3 & 3 & 3 & 2 & 4 & 3.0 & 0.63 \\ 
			\hline
			
			\multirow{5}{*}{Phi-3 Mini} 
			& Zero-Shot & 1 & 4 & 4 & 4 & 2 & 3.0 & 1.22 \\ \cline{2-9}
			& Few-Shot & 2 & 4 & 4 & 4 & 3 & 3.4 & 0.89 \\ \cline{2-9}
			& Chain-of-Thought & 1 & 3 & 4 & 4 & 3 & 3.0 & 1.10 \\ \cline{2-9}
			& Constraint-based & 1 & 3 & 4 & 5 & 2 & 3.0 & 1.22 \\ \cline{2-9}
			& Role-based & 2 & 4 & 3 & 4 & 3 & 3.2 & 0.75 \\ 
			\hline
			\textbf{Average} & -- & 2.6 & 4.0 & 3.0 & 2.3 & 3.8 & 3.1 & 0.8 \\ 
			\hline
		\end{tabular}
	}
\end{table*}

}

\subsection{Discussion}
\label{sec:discussion-1}
Lightweight LLMs show promising potential for analysing user feedback in RE. Although more efficient and transparent than larger models, they are not yet suitable for reliable industrial use without further customization.

\noindent\textbf{A) Feedback Classification.} In feedback classification, Llama 3 and Mistral achieved F1-scores around 0.74 for identifying user request types. This shows that lightweight models can reliably detect explicit requests such as feature suggestions or bug reports. Their performance in NFR classification was weaker, with the best F1-score reaching 0.55. This result aligns with prior studies where models also struggled to capture implicit qualities like usability or reliability~\cite{lu2017automatic}. The limitation likely stems from two factors. Lightweight LLMs find it difficult to interpret subtle, context-dependent expressions. They also lack sufficient exposure to RE-specific terminology and training data. Structured prompting with few-shot or reasoning examples led to only minor improvements. These findings suggest that contextual examples and reasoning cues can enhance understanding. However, overall accuracy remains moderate. In practice, this may produce noisy classifications that mislead analysts. As a result, important feedback can be missed, while irrelevant comments may be misclassified as critical requirements.

\noindent\textbf{B) Requirements Specification Generation}. Lightweight LLMs produced requirement specification that were generally clear, coherent, and complete. They expressed user feedback readably but often failed to follow formal structures. Their outputs were sometimes verbose and required editing for conciseness and compliance with standards. Although the generated text was fluent, the models occasionally fabricated requirements, adding information absent from the original feedback. As a result, the content appeared plausible but not always accurate or grounded. In practice, these limitations mean lightweight LLMs can assist analysts by drafting initial requirement statements or summarising feedback. However, they still require human review to ensure accuracy and proper structure. With further refinement and adaptation, they could serve as drafting and documentation aids rather than autonomous specification generators.

\noindent\textbf{C) Implications for Requirements Engineering.} Lightweight LLMs can support RE tasks such as feedback filtering, classification, and initial specification drafting. However, they cannot yet replace human analysts. Their moderate precision and recall make them unreliable for use without supervision. These weaknesses may lead to missed insights or false positives that increase review effort.
The generated outputs are clear and coherent but often lack factual grounding and formal consistency. This limits their value for downstream RE tasks, e.g., validation, and traceability. In several tasks, their performance is similar to earlier ML approaches~\cite{Dabrowski2022,lu2017automatic}. Larger models alone do not guarantee better outcomes in RE. Future work should adapt models to RE contexts using prompt design, fine-tuning, and retrieval-based approaches. These methods can improve accuracy and help lightweight LLMs better support early RE tasks.

\subsection{Threats to Validity}
\label{sec:threats-1}
\noindent\textbf{Internal Validity.}
The main threat lies in the manual evaluation of generated requirement specification by a single evaluator, \JD{without validating its reliability}. This introduces potential subjectivity. To mitigate this, we applied a systematic rubric with clear criteria and examples of both high- and low-quality specification. The rubric was used consistently across all outputs. In addition, prompts and model parameters were standardised to ensure uniform conditions.

\noindent\textbf{External Validity.}
We used two publicly available datasets from different domains to reduce domain bias. Their diverse vocabulary helps generalisability, though an app review represents only one feedback type. Results may not generalise to industrial datasets or other contexts such as issue trackers. Moreover, the study focused on lightweight LLMs; larger models may yield different results.

\noindent\textbf{Construct Validity.}
We employed standard precision, recall, and F1-score metrics for classification and a rubric assessing completeness, consistency, and correctness for generation. As qualitative evaluation relied on one evaluator, some interpretation bias may persist. Prompt phrasing may also influence results; this was mitigated by systematically applying established prompting strategies.

\noindent\textbf{Conclusion Validity.}
All models were tested under identical settings and prompts. However, the limited sample size for specification generation lowers statistical power, and no inferential tests were applied, making results exploratory. While precision and recall were key metrics, tasks may favour one over the other, and F1 may not always be optimal~\cite{Berry2022}. \JD{Our aim was to assess overall practical effectiveness rather than examine differences across RE-specific tasks.}

\section{Study II: Requirements Across Project Artefacts}\label{sec:study-2}
In the second study, we investigate whether frontier LLMs can support requirements traceability across heterogeneous software project artefacts. Unlike the controlled experiment presented in Study I (Sect.~\ref{sec:study-1}), Study II is designed as an exploratory industrial case study based on observations from a research visit at Huawei Research Centre. We evaluate two frontier LLMs on two complementary RE tasks: traceability link identification and traceability explanation generation~\cite{zadenoori2025largelanguagemodelsllms}. We first introduce the industrial study context, followed by the empirical study design, experimental findings, and their implications for RE practice.

\subsection{Industrial Study Context}
Our second study focuses on analysing requirements-related information distributed across software project artefacts. Specifically, we consider \textit{project goals}\footnotemark[\getrefnumber{fn:project_rust_goal}], \textit{GitHub tracking issues}\footnotemark[\getrefnumber{fn:project_rust_issue}], and \textit{Zulip messages}\footnotemark[\getrefnumber{fn:project_rust_zulip}] from the Rust open-source ecosystem\footnotemark[\getrefnumber{fn:project_rust}]. These artefacts capture project objectives, implementation activities, technical decisions, and project evolution. As requirements-related information is distributed across multiple artefacts, reconstructing requirement evolution is labour-intensive, motivating automated support for traceability and rationale analysis.

A \textit{project goal} defines a high-level development objective, its motivation, and expected outcomes. A \textit{tracking issue} describes the implementation of a project goal, including implementation tasks, milestones, dependencies, and progress updates. A \textit{Zulip message} captures developer discussions on project goals or tracking issues, including technical reasoning, design decisions, and implementation details. A \textit{traceability link} links a project goal or tracking issue with a related Zulip message describing the same development objective or implementation activity.

The study evaluates two RE tasks. The first, \textit{traceability link identification}, identifies Zulip messages related to a given project goal or tracking issue. The second, \textit{traceability explanation generation}, produces an evidence-grounded explanation describing the identified traceability link and its supporting evidence.

For example, consider the project goal \textit{``Improve compiler support for Rust-for-Linux by stabilizing the required language features.''} A related Zulip message states: \textit{``The remaining compiler changes have been merged, and Rust-for-Linux now builds successfully on nightly.''} The first task identifies the message as related to the project goal. The second generates the explanation: \textit{``The message provides evidence that the remaining compiler changes have been completed, demonstrating implementation progress towards the Rust-for-Linux compiler support goal.''}

\subsection{Empirical Study Design}
\label{sec:design-2}
This study evaluates the ability of frontier LLMs to identify traceability links across heterogeneous software project artefacts and generate evidence-grounded explanations. Unlike Study I (see Sect.~\ref{sec:study-1}), it adopts an industrial case study to evaluate readily available frontier LLMs in a realistic setting.

\subsubsection{Research Questions}
\label{sec:RQs-2}

The goal of this study is to evaluate LLMs in supporting traceability analysis across distributed software project artefacts. We specifically focus on two research questions:

\begin{itemize}
	\item \textbf{RQ4}: How well do LLMs identify traceability links across project artefacts?
	\item \textbf{RQ5}: How well do LLMs generate traceability explanations?
\end{itemize}

In RQ4, we assess the ability of LLMs to identify traceability links between Rust Project Goals or GitHub tracking issues and related Zulip discussions. RQ5 evaluates the quality of the explanations generated for the identified traceability links. All generated links and explanations are manually evaluated using predefined assessment criteria (see Sect.~\ref{sec:metrics2}). For RQ4, retrieved links are validated by human assessors and analysed using Precision@N and NDCG@N. For RQ5 explanations are assessed through human judgment based on predefined quality criteria.

\subsubsection{Dataset}
\label{sec:dataset-2}

We collected a new dataset from the Rust open-source ecosystem to evaluate traceability identification and explanation generation across software project artefacts\footnotemark[\getrefnumber{fn:project_rust}]. The Rust ecosystem was selected because our industrial partner actively contributes to its development, making cross-artefact traceability a practically relevant engineering challenge. The dataset comprises nine traceability cases from the 2024--2026 Rust Project Goals initiative\footnotemark[\getrefnumber{fn:project_rust_goal}]. The cases were selected to ensure complete traceability scenarios, diversity of technical domains, and coverage of multiple roadmap years (2024--2026). Each case comprises a Rust Project Goal, its corresponding GitHub tracking issue\footnotemark[\getrefnumber{fn:project_rust_issue}], and the associated Zulip discussion\footnotemark[\getrefnumber{fn:project_rust_zulip}]:

\begin{itemize}
	\item \textbf{Project Goals} describe strategic objectives, motivations, and expected outcomes.
	\item \textbf{GitHub Tracking Issues} capture implementation planning, task management, dependencies, and progress updates.
	\item \textbf{Zulip Discussions} contain technical reasoning, design decisions, coordination activities, and implementation details.
\end{itemize}

We retained only project goals with linked GitHub tracking issues and related Zulip discussions. The associated Zulip discussions range from 384 to 39,343 messages, comprising 80,642 messages in total. Table~\ref{tab:dataset-overview} summarises the collected traceability cases, including their year, topic, description, and the number of associated Zulip messages. The cases span compiler engineering, tooling, language design, verification, infrastructure, and project governance. The dataset is available in the replication package~\cite{supplementary_materials}.

\begin{table*}[t]
	\centering
	\caption{Overview of the Rust traceability dataset used in Study II. Each traceability case comprises a Rust Project Goal, its associated GitHub tracking issue, and the corresponding Zulip discussion. The table reports the project year, topic, description, and the number of associated Zulip messages.}
	\small
	\begin{tabular}{|c|c|p{3cm}|p{4.3cm}|c|}
		\hline
		\textbf{Case} & \textbf{Year} & \textbf{Topic} & \textbf{Description} & \textbf{No. Msgs} \\
		\hline
		1 & 2024 & MIR Formality & Formal type system model & 1,036 \\
		2 & 2024 & Polonius & Scalable borrow checker & 14,908 \\
		3 & 2024 & Rust for Linux & Stable Linux kernel support & 1,808 \\
		4 & 2025 & Async Rust & Async language improvements & 39,343 \\
		5 & 2025 & Parallel Front End & Parallel compiler frontend & 6,045 \\
		6 & 2025 & StableMIR & Stable compiler API & 2,982 \\
		7 & 2026 & Build-std & Standard library builds & 384 \\
		8 & 2026 & Type Documentation & Type system documentation & 13,047 \\
		9 & 2026 & User Research & User research team & 1,089 \\
		\hline
		\textbf{Total} & -- & -- & -- & \textbf{80,642} \\
		\hline
	\end{tabular}
	\label{tab:dataset-overview}
\end{table*}

\subsubsection{Pilot Study}

We first conducted a pilot study to assess the feasibility of LLM-based traceability identification and explanation generation across distributed software project artefacts. The pilot used two Rust project goals together with their associated GitHub issues and Zulip discussions. It refined the prompting strategy, artefact selection procedure, and manual validation protocol for the full study.

We initially evaluated the best-performing lightweight LLMs from Study~I (see Sect.~\ref{sec:background}), namely Mistral and Llama. Both models performed well on feedback-driven RE tasks. However, their limited context windows restricted reasoning across heterogeneous software project artefacts. Local execution was also computationally demanding. We therefore selected frontier LLMs for the main study.

The pilot confirmed the feasibility of LLM-supported traceability analysis. Evidence-oriented and chain-of-thought prompting produced more technically grounded explanations than simple retrieval prompts. Finer-grained technical artefacts improved retrieval precision. In contrast, retrieving entire discussion threads introduced contextual noise through semantically related but unsupported discussions.

The pilot also showed that Goal--Zulip and Issue--Zulip retrieval are complementary tasks with different levels of ambiguity. We therefore evaluated them separately. Finally, the pilot confirmed the need for manual validation. Plausible-looking links often required verification against the surrounding discussions and Rust documentation. These findings informed the design of the full study.

\subsubsection{Evaluation Metrics and Criteria}
\label{sec:metrics2}

We applied both quantitative and qualitative methods. Retrieval metrics were used for RQ4, as traceability identification is formulated as a retrieval task. RQ5 was evaluated through criteria-based assessment using predefined quality attributes.

\paragraph{Evaluation Metrics for Traceability Identification (RQ4).}

For RQ4, we evaluate the ability of LLMs to identify traceability links between Project Goals or GitHub tracking issues and related Zulip discussions. Retrieved links are manually validated by human assessors. We report Precision@5 and Normalized Discounted Cumulative Gain (NDCG@5)~\cite{manning2008introduction}. We selected a cutoff of five because manual validation of retrieved links is time-intensive, while the highest-ranked candidates are the most relevant in practical traceability analysis. Precision@5 measures the proportion of validated links among the top five retrieved candidates, while NDCG@5 evaluates the quality of their ranking. Goal--Zulip and Issue--Zulip retrieval are evaluated separately. Recall is not reported because no complete ground truth of traceability links is available for the selected artefacts~\cite{manning2008introduction}. This is common in large-scale information retrieval tasks, where exhaustive identification of all relevant links is infeasible~\cite{Dabrowski2023}.

\paragraph{Evaluation Criteria for Explanation Generation (RQ5).}

For RQ5, traceability explanations are evaluated qualitatively using two criteria: \textit{Correctness} and \textit{Evidence Grounding}. Table~\ref{tab:explanation_rubric} summarises the evaluation rubric. Both criteria assess whether explanations correctly justify the identified traceability links and are supported by evidence from the referenced project artefacts. Each criterion is rated on a five-point Likert scale (Table~\ref{tab:explanation_scale})~\cite{kuckartz2014qualitative}. We report the mean scores across all validated traceability links.

\begin{table}[t]
	\centering
	\caption{Evaluation criteria and scoring used to assess generated traceability explanations (RQ5).}
	\label{tab:explanation_rubric}
	\small
	\begin{tabular}{|p{3.5cm}|p{9cm}|}
		\hline
		\textbf{Criterion} & \textbf{Description and Scoring} \\
		\hline
		
		\multirow{3}{*}{\centering\textbf{Correctness}} &
		\textbf{Description:} Evaluates whether the explanation accurately justifies the identified traceability link and correctly describes the relationship between the linked artefacts. \\
		\cline{2-2}
		&
		\textbf{Scoring:} Rated on a 1--5 Likert scale, where higher scores indicate greater correctness. \\
		\hline
		
		\multirow{5}{*}{\centering\textbf{Evidence Grounding}} &
		\textbf{Description:} Evaluates whether the explanation is supported by information contained in the referenced artefacts without introducing unsupported claims. \\
		\cline{2-2}
		&
		\textbf{Scoring:} Rated on a 1--5 Likert scale, where higher scores indicate stronger evidence grounding. \\
		\hline
		
	\end{tabular}
\end{table}

\begin{table*}[t]
	\centering
	\caption{Interpretation of the five-point Likert scale used to evaluate explanation quality.}
	\label{tab:explanation_scale}
	\small
	\begin{tabular}{|c|p{5.5cm}|p{5.5cm}|}
		\hline
		\textbf{Score} & \textbf{Correctness} & \textbf{Evidence Grounding} \\
		\hline
		
		1 &
		Explanation is incorrect, misleading, or contradicts the referenced artefacts. &
		Evidence is missing, irrelevant, or does not support the identified traceability link. \\
		\hline
		
		2 &
		Explanation contains major inaccuracies or unsupported reasoning. &
		Evidence is weak, vague, or only loosely related to the identified link. \\
		\hline
		
		3 &
		Explanation is partially correct but incomplete or insufficiently justified. &
		Evidence is partially relevant but provides only limited support for the identified link. \\
		\hline
		
		4 &
		Explanation is mostly accurate and clearly describes the relationship between the artefacts. &
		Evidence is relevant and provides adequate support for the identified link. \\
		\hline
		
		5 &
		Explanation is accurate, precise, and fully consistent with the referenced artefacts. &
		Evidence is highly relevant, specific, verifiable, and fully supports the identified link. \\
		\hline
		
	\end{tabular}
\end{table*}

\subsubsection{Experimental Setup and Procedure}
\label{sec:setup-2}

We now describe the computational setup, prompting strategies, and evaluation procedures used to address RQ4 and RQ5.

\paragraph{Computational Setup}
\label{sec:comp-setup-2}

We evaluated GPT-5.5 and DeepSeek-R1 using their latest publicly available versions through the official provider-hosted interfaces. GPT-5.5 was selected as a representative frontier LLM, whereas DeepSeek-R1 represents a leading open-weight reasoning model of interest to our industrial partner. Experiments were managed locally on a MacBook Pro (Apple M2 Pro, 32~GB RAM), while all model inference was performed on the providers' infrastructure.

Unlike Study~I (Sect.~\ref{sec:study-1}), each experimental configuration was executed once. As this study aims to evaluate practical capability rather than performance variability, repeated executions were not required. The experimental design comprised 72 executions, covering all combinations of two frontier LLMs, two prompting strategies, two document types, and nine project topics. Response times ranged from approximately 30 seconds to 2 minutes per query, supporting interactive traceability analysis.

\paragraph{Prompting Strategies}
\label{sec:prompts-2}

We adopted two prompting strategies introduced in Section~\ref{sec:background}: zero-shot prompting and chain-of-thought (CoT) prompting. These strategies were selected because they provide two complementary reasoning settings while keeping the prompting design simple. Zero-shot prompting serves as a baseline, whereas CoT explicitly supports the multi-step reasoning required for traceability analysis. More complex strategies (e.g., few-shot or role-based prompting) were intentionally excluded to avoid introducing additional prompt-specific effects, as the objective of this study was to evaluate the reasoning capabilities of frontier LLMs rather than optimise prompt design.

Prompt development followed an iterative process during the pilot study. A small subset of project artefacts, excluded from the final evaluation, was used to refine task instructions, reduce ambiguities, and improve evidence grounding. The prompts were revised until they produced consistent outputs across both evaluated models. Once finalised, they remained unchanged throughout the experiment to ensure a fair comparison. Table~\ref{tab:prompt_examples} presents simplified examples of the zero-shot and CoT prompts. The complete prompt templates are available in the replication package~\cite{supplementary_materials}.

\begin{table}[t]
	\centering
	\caption{Prompting strategies and simplified examples used in the study.}
	\label{tab:prompt_examples}
	\small
	\begin{tabular}{|p{3.5cm}|p{8.5cm}|}
		\hline
		\textbf{Prompting Strategy} & \textbf{Simplified Example} \\
		\hline
		\textbf{Zero-shot} &
		Identify the most relevant Zulip discussion fragments related to the following Rust project goal. Return the top-5 candidate links together with a short explanation. \\
		\hline
		\textbf{Chain-of-thought} &
		Analyse the following Rust project goal and Zulip discussion fragments step by step. Explain why the artefacts may be related, identify supporting evidence, and return the top-5 most relevant traceability links with explanations. \\
		\hline
	\end{tabular}
\end{table}

\paragraph{Evaluation Procedures}
\label{sec:eval-proc-2}

We applied quantitative evaluation for traceability identification (RQ4) and qualitative evaluation for explanation generation (RQ5).\newline

\noindent\textit{Evaluation Procedure for Traceability Identification (RQ4).}
For each Project Goal or GitHub tracking issue, we provided the corresponding candidate Zulip discussions to the LLM. Under each prompting strategy, the model identified and ranked the five most relevant discussion fragments, extracted supporting evidence, and generated a short explanation for each identified link. Retrieved links were manually validated using predefined assessment guidelines. A traceability link was considered valid when the linked artefacts referred to the same project objective, implementation activity, technical concern, dependency, or design discussion. Precision@5 and NDCG@5 were then computed from the validated rankings.\newline

\noindent\textit{Evaluation Procedure for Explanation Generation (RQ5).}
For RQ5, we manually evaluated the explanations associated with the validated traceability links using the criteria described in Sect.~\ref{sec:metrics}. Correctness and Evidence Grounding were assessed independently for each explanation using a five-point Likert scale. We report the mean and median scores across all validated traceability links.\newline

To assess the reliability of the manual evaluation, a second author independently reviewed a randomly selected 10\% subset of the evaluated cases. Cohen's Kappa reached 0.83 for traceability link validation~\cite{manning2008introduction}, indicating almost perfect agreement, whereas Quadratic Weighted Cohen's Kappa reached 0.71 for explanation quality ratings, indicating substantial agreement~\cite{cohen1968weighted}.

\subsection{Results}
\label{sec:results-2}
\noindent\textbf{RQ4: How well do LLMs identify traceability links across project artefacts?}

\noindent To answer RQ4, we evaluated how well frontier LLMs identify traceability links between project goals, GitHub tracking issues, and Zulip discussions under two prompting strategies: zero-shot and chain-of-thought. Table~\ref{tab:llm-retrieval-performance} reports Precision@5 (P@5) and Normalized Discounted Cumulative Gain (NDCG@5) for Goal--Zulip and Issue--Zulip traceability link identification, with the best results highlighted in bold. For Goal--Zulip traceability link identification, chain-of-thought prompting consistently improved performance. The average P@5 increased from 0.59 to 0.73, while the average NDCG@5 increased from 0.87 to 0.90. ChatGPT achieved the highest NDCG@5 (0.94), whereas both models achieved the highest P@5 (0.73) under chain-of-thought prompting. For Issue--Zulip traceability link identification, ChatGPT consistently outperformed DeepSeek. Under chain-of-thought prompting, it achieved the highest P@5 (0.91) and NDCG@5 (0.98). In contrast, DeepSeek performed best under zero-shot prompting, with performance decreasing under chain-of-thought prompting. Overall, the models performed better on Issue--Zulip than Goal--Zulip traceability link identification, achieving a higher average P@5 (0.77 versus 0.73).

\rqanswerfour{Frontier LLMs achieved moderate-to-high performance for traceability link identification. Chain-of-thought prompting improved Goal--Zulip links, while ChatGPT performed best overall (P@5=0.91, NDCG@5=0.98).}

\begin{table*}[h]
	\centering
	\caption{Performance of frontier LLMs on traceability link identification (RQ4). P@5 and NDCG@5 are reported for Goal--Zulip and Issue--Zulip links under zero-shot and chain-of-thought prompting. Best results are shown in bold.}
	\label{tab:llm-retrieval-performance}
	\setlength{\tabcolsep}{6pt}
	\renewcommand{\arraystretch}{1.2}
	\scalebox{0.8}{
		\begin{tabular}{|l|c|c|c|c|c|c|c|c|}
			\hline
			\multirow{3}{*}{\textbf{Model}}
			& \multicolumn{4}{c|}{\textbf{Goal--Zulip Link}}
			& \multicolumn{4}{c|}{\textbf{Issue--Zulip Link}} \\
			\cline{2-9}
			& \multicolumn{2}{c|}{\textbf{Zero-Shot}}
			& \multicolumn{2}{c|}{\textbf{Chain-of-Thought}}
			& \multicolumn{2}{c|}{\textbf{Zero-Shot}}
			& \multicolumn{2}{c|}{\textbf{Chain-of-Thought}} \\
			\cline{2-9}
			& \textbf{P@5} & \textbf{NDCG@5}
			& \textbf{P@5} & \textbf{NDCG@5}
			& \textbf{P@5} & \textbf{NDCG@5}
			& \textbf{P@5} & \textbf{NDCG@5} \\
			\hline
			ChatGPT  & 0.64 & \textbf{0.94} & \textbf{0.73} & \textbf{0.91} & \textbf{0.78} & \textbf{0.94} & \textbf{0.91} & \textbf{0.98} \\
			DeepSeek & 0.53 & 0.80          & \textbf{0.73} & 0.89          & 0.67          & \textbf{0.94} & 0.62          & 0.89 \\
			\hline
			\textbf{Average} & 0.59 & 0.87 & 0.73 & 0.90 & 0.73 & 0.94 & 0.77 & 0.94 \\
			\hline
		\end{tabular}
	}
\end{table*}

\noindent\textbf{RQ5: How well do LLMs generate traceability explanations?}

\noindent To answer RQ5, each model was evaluated on two criteria: correctness (CR) and evidence grounding (EG), as defined in Table~\ref{tab:explanation_rubric}. Table~\ref{tab:llm-summary-performance} reports results across models, prompting strategies, and traceability link types. Overall, the models produced high-quality traceability explanations. Mean scores ranged from 3.94 to 4.35, with average scores of 4.15 for Goal--Zulip explanations and 4.21 for Issue--Zulip explanations. DeepSeek with chain-of-thought prompting achieved the highest mean score for Goal--Zulip explanations (4.27), showing strong correctness and evidence grounding (CR=4.21; EG=4.33), while ChatGPT achieved the highest correctness score (CR=4.23). ChatGPT with zero-shot prompting achieved the highest mean score for Issue--Zulip explanations (4.35), demonstrating the best correctness and evidence grounding (CR=4.35; EG=4.34). Across prompting strategies, chain-of-thought prompting improved Goal--Zulip explanation quality for both models, whereas DeepSeek showed only modest improvements for Issue--Zulip explanations. Overall, Issue--Zulip explanations received slightly higher scores than Goal--Zulip explanations.

\rqanswerfive{Frontier LLMs produced high-quality traceability explanations. Chain-of-thought prompting improved Goal--Zulip explanations; ChatGPT achieved the best performance for Issue--Zulip explanations (Mean = 4.35).}

\begin{table*}[h]
	\centering
	\caption{Performance of frontier LLMs on traceability explanation generation (RQ5).
		CR -- Correctness; EG -- Evidence Grounding.
		Scores are on a 1--5 scale (higher = better).
		Mean denotes the average of CR and EG.
		\textbf{Bold} indicates the best score for each evaluation criterion.}
	\label{tab:llm-summary-performance}
	\setlength{\tabcolsep}{5pt}
	\renewcommand{\arraystretch}{1.2}
	\scalebox{0.8}{
		\begin{tabular}{|l|l|c|c|c|c|c|c|}
			\hline
			\multirow{2}{*}{\textbf{Model}}&\multirow{2}{*}{ 	\textbf{Prompt Strategy}} &
			\multicolumn{3}{c|}{\textbf{Goal--Zulip}} &
			\multicolumn{3}{c|}{\textbf{Issue--Zulip}} \\
			\cline{3-8}
			& &
			\textbf{CR (1--5)} &
			\textbf{EG (1--5)} &
			\textbf{Mean} &
			\textbf{CR (1--5)} &
			\textbf{EG (1--5)} &
			\textbf{Mean} \\
			\hline
			
			\multirow{2}{*}{ChatGPT}
			& Zero-Shot
			& 3.89 & 3.99 & 3.94
			& \textbf{4.35} & \textbf{4.34} & \textbf{4.35} \\
			\cline{2-8}
			& Chain-of-Thought
			& \textbf{4.23} & 4.18 & 4.21
			& 4.09 & 4.12 & 4.11 \\
			\hline
			
			\multirow{2}{*}{DeepSeek}
			& Zero-Shot
			& 4.14 & 4.19 & 4.17
			& 4.17 & 4.15 & 4.16 \\
			\cline{2-8}
			& Chain-of-Thought
			& 4.21 & \textbf{4.33} & \textbf{4.27}
			& 4.23 & 4.20 & 4.22 \\
			\hline
			
			\textbf{Average}
			& --
			& 4.12 & 4.17 & 4.15
			& 4.21 & 4.20 & 4.21 \\
			\hline
		\end{tabular}
	}
\end{table*}

\subsection{Discussion}
\label{sec:discussion-2}
Frontier LLMs show promising potential for supporting requirements traceability across software project artefacts. Although they identify traceability links and generate explanations, human verification remains necessary for reliable industrial use.

\noindent\textbf{A) Traceability Link Identification.}
Frontier LLMs achieved moderate-to-high performance in identifying traceability links. Performance was consistently higher for Issue--Zulip than Goal--Zulip links. This is expected because roadmap goals describe high-level strategic objectives, whereas GitHub issues refine these objectives into concrete implementation tasks. Zulip discussions typically focus on these implementation activities, making Issue--Zulip links semantically more direct. In contrast, Goal--Zulip links span different levels of abstraction and therefore require greater semantic reasoning. Chain-of-thought prompting particularly improved Goal--Zulip link identification, indicating that explicit reasoning helps bridge the gap between strategic objectives and implementation discussions. These findings suggest that frontier LLMs can reduce the effort of recovering traceability links across heterogeneous software project artefacts. Nevertheless, human verification remains necessary when traceability information supports downstream activities such as impact analysis or release planning.

\noindent\textbf{B) Traceability Explanation Generation.}
Frontier LLMs generated explanations that were generally correct and evidence-grounded. They successfully explained why project artefacts were related and summarized the underlying design rationale. Goal--Zulip explanations benefited more from chain-of-thought prompting than Issue--Zulip explanations. This is expected as project goals capture high-level objectives, while the supporting evidence is distributed across multiple implementation discussions. Explaining these relationships therefore requires reasoning across different levels of abstraction. In contrast, GitHub issues and Zulip discussions are semantically closer and typically provide sufficient contextual evidence without additional reasoning. Automatically generated explanations can help developers understand why project artefacts are related and provide useful context during repository exploration. However, they should be treated as supporting evidence rather than authoritative rationale, particularly when decisions depend on interpretation of project history.

\noindent\textbf{C) Implications for Requirements Engineering.}
Frontier LLMs achieved sufficient performance to support requirements traceability across heterogeneous software project artefacts. Both traceability identification and explanation generation can help practitioners reconstruct relationships between strategic project goals, GitHub issues, and developer discussions with substantially less manual effort. This can improve design rationale understanding, requirements impact analysis, and developer onboarding. The results also highlight an important trade-off for industrial adoption. Frontier LLMs rely on commercial providers, raising concerns about vendor lock-in, privacy, and operational costs. However, our pilot study found that current off-the-shelf lightweight LLMs are not yet practical. Their limited context windows and slow local execution hinder analysis of large software repositories. Moreover, requirements traceability is only one of many RE activities. Organisations are therefore unlikely to customise and maintain lightweight LLMs solely for this task. General-purpose frontier LLMs that support multiple RE/SE activities may currently offer the more practical solution. Nevertheless, frontier LLMs should augment rather than replace engineers. Incorrect traceability links or explanations may otherwise propagate errors to engineering activities such as requirements impact analysis and release planning.

\subsection{Threats to Validity}
\label{sec:threats-2}
\noindent\textbf{Internal Validity.} The main threat lies in the manual validation of traceability links and explanation quality. Assessing whether a retrieved discussion genuinely supports a project goal or issue often requires interpretation of technical context and Rust-specific concepts. To mitigate this threat, we applied predefined validation criteria and evaluation guidelines throughout the study. A second evaluator independently assessed a randomly selected subset of the results. Substantial inter-rater agreement supported the reliability of the manual evaluation. When ambiguities arose, official Rust documentation and tutorials were consulted. Nevertheless, some interpretation bias may remain.

\noindent\textbf{External Validity.} A threat to external validity concerns the generalisability of the results. The study was conducted using a single case study based on the Rust ecosystem. The project was selected because it provides a rare combination of publicly available project goals, tracking issues, and developer discussions, enabling transparent and reproducible evaluation. Comparable industrial datasets are rarely publicly available due to confidentiality constraints. While Rust is a large and industrially relevant open-source project, it represents only one development context. Other open-source or industrial projects may follow different development practices and rely on different artefacts. We do not claim direct generalisability. Rather, the goal of this study is to provide an in-depth illustration of applying LLM-supported traceability analysis in a realistic open-source setting.

\noindent\textbf{Construct Validity.} Another threat concerns whether the applied measures adequately capture the intended requirements engineering constructs. Concepts such as traceability links, rationale, dependencies, and explanation quality originate from established requirements engineering and traceability research, whereas the analysed artefacts are informal and community-driven. As a result, some relationships may only partially reflect these constructs or oversimplify them. To mitigate this threat, we grounded the evaluation criteria in established traceability and information-retrieval measures, and assessed explanation quality using predefined dimensions of correctness and evidence grounding. Nevertheless, some constructs may remain underspecified due to the informal nature of the source artefacts.

\noindent\textbf{Conclusion Validity.} A threat to conclusion validity concerns the strength and reliability of the conclusions. The study does not aim to establish causal relationships or benchmark model performance. Instead, it reports observations from applying off-the-shelf LLMs in a realistic open-source software setting. The number of analysed artefacts is limited by the effort required for manual validation. Furthermore, no complete ground truth exists for the selected artefacts. These factors restrict the strength of the conclusions. The findings should therefore be interpreted as exploratory insights into the practical usefulness of LLM-supported traceability analysis rather than definitive evidence of effectiveness.

\section{Cross-study Discussion}\label{sec:cross-study-discussion}
Although the two studies investigated different RE activities, artefacts, models, and evaluation settings, they provide complementary evidence on the capabilities and limitations of LLMs for RE. This section synthesises the findings and discusses their broader implications for RE research and practice.

\subsection{Cross-Task Perspective on LLM Performance in RE}
The two empirical studies provide a broader perspective on LLM capabilities across five complementary RE activities: requirements classification, requirements specification, traceability identification, and traceability explanation generation. Table~\ref{tab:cross-task-summary} synthesises the empirical findings by summarising the best-performing model, prompting strategy, and overall performance achieved for each RE task. The synthesis shows that no single model or prompting strategy consistently achieved the best performance across all RE activities. Instead, LLM effectiveness was strongly task-dependent, with different models proving more suitable for different RE tasks. Performance ranged from moderate to high depending on the reasoning demands, artefact characteristics, and expected outputs of the target activity. Accordingly, the findings should be interpreted as evidence of the practical readiness of current LLMs for supporting specific RE activities rather than as a ranking of models or prompting strategies. From an RE perspective, these findings suggest that RE/SE practitioners should adopt LLMs selectively rather than uniformly across the requirements lifecycle. Model selection should be guided by the characteristics and reasoning demands of the target RE task rather than by overall model rankings. Human expertise nevertheless remains essential wherever engineering judgement and validation are required.

\begin{table}[t]
	\centering
	\caption{Cross-task empirical understanding of LLM performance across Requirements Engineering (RE) activities.}
	\label{tab:cross-task-summary}
	\small
	\begin{tabular}{|l|l|l|c|}
		\hline
		\textbf{RE Task} & \textbf{Best Model} & \textbf{Best Prompt} & \textbf{Performance} \\
		\hline
		NFR Classification & Gemma & Chain-of-Thought & Moderate \\
		\hline
		User Request Classification & Llama 3 & Few-shot & High \\
		\hline
		Requirements Specification & Llama 3 / Mistral & CoT / Few-shot & Moderate \\
		\hline
		Traceability Identification & GPT-5.5 & Chain-of-Thought & High \\
		\hline
		Traceability Explanation & GPT-5.5 & Chain-of-Thought & Moderate--High \\
		\hline
	\end{tabular}
\end{table}

\subsection{Prompting Matters, but Depends on the RE Task}
Across both studies, prompting strategy consistently influenced LLM performance, although the most effective strategy varied according between RE activities. As summarised in Table~\ref{tab:cross-task-summary}, Chain-of-Thought prompting generally benefited reasoning-intensive tasks, including NFR classification, traceability identification, and traceability explanation generation. In contrast, few-shot prompting proved most effective for user request classification by providing representative examples that reduced ambiguity. These findings suggest that prompt engineering should be viewed as RE task-specific rather than universally applicable. Consequently, prompting strategies should be selected according to the reasoning demands and expected outputs of the target RE task rather than applied uniformly across the RE lifecycle.

\subsection{Human Oversight Remains Essential}
Although both studies demonstrate the potential of LLMs to support multiple RE activities, they also show that human oversight remains essential. The level of supervision required, however, depends on the complexity of the target RE task.
Tasks involving relatively well-defined decisions, such as user request classification and traceability identification, generally produced reliable outputs that can reduce manual effort, although verification remains necessary.
In contrast, tasks requiring more complex reasoning, synthesising information across multiple artefacts, or generating requirements-related artefacts were more susceptible to unsupported, incomplete, or insufficiently grounded outputs. These differences can largely be explained by the reasoning demands of individual RE activities. As tasks require synthesising information across multiple artefacts, preserving semantic consistency, or generating new textual artefacts, the likelihood of grounding and consistency issues increases. Consequently, LLMs should be viewed as decision-support tools that augment rather than replace RE practitioners.

\subsection{Practical Implications for Requirements Engineering}
The findings have several practical implications for RE practice. First, LLM adoption should be driven by the characteristics and reasoning demands of the target RE task rather than by model popularity or benchmark rankings. Lightweight open-source models performed well on short-context tasks, such as analysing user feedback. They also offer advantages in local deployment, reproducibility, and computational efficiency. However, our pilot experimentation showed that these models were impractical for traceability analysis across heterogeneous software project artefacts because of their limited context windows, weaker reasoning capabilities, and substantially longer execution times. Consequently, frontier models were adopted for Study~II to support reasoning across long and heterogeneous artefacts. Second, model selection and prompt design should be considered jointly. Although prompting consistently influenced performance, the most effective strategy depended on the reasoning demands of the target RE task rather than on the model alone. Finally, LLM outputs should be treated as decision-support artefacts rather than final engineering products, regardless of the model family. Requirements classifications, traceability links, generated specifications, and explanations should all be validated before being incorporated into SE practice. Overall, our findings indicate that the successful adoption of LLMs in RE depends less on identifying a universally superior model than on selecting models and prompting strategies that match the characteristics and reasoning demands of the target RE task.

\subsection{Future Research Directions}
The cross-task perspective presented in this work also highlights several opportunities for future research. First, improving factual grounding remains an important challenge, particularly for requirements specification generation and other generation-oriented RE activities. Second, future work should investigate hybrid approaches combining prompting with retrieval-augmented generation, repository-aware retrieval, or fine-tuning on RE-specific corpora to improve factual consistency and domain adaptation. More broadly, evaluating LLMs across multiple RE activities using common datasets, evaluation protocols, and replication packages would support more systematic cross-task comparisons and strengthen cumulative empirical evidence on the role of LLMs throughout the RE lifecycle. Rather than evaluating individual RE tasks in isolation, future studies should increasingly investigate how LLMs support complementary activities across the RE lifecycle. Future studies should therefore move beyond evaluating individual RE tasks in isolation and instead investigate how LLMs support complementary activities across the RE lifecycle.

\section{Conclusion}\label{sec:conclusion}
Requirements-related information is increasingly distributed across heterogeneous software project artefacts, making many RE activities labour-intensive and difficult to scale. Large LLMs offer new opportunities to support these activities~\cite{cheng2024generative}. This paper addressed the fragmented empirical evidence on LLMs for RE through the first cross-task empirical evaluation spanning five representative RE activities. By integrating evidence from two complementary empirical studies, it provides a broader perspective on the capabilities and limitations of current LLMs across different RE contexts.

The findings demonstrate that LLM effectiveness is strongly RE task-dependent. Current LLMs can effectively support several RE activities, but their performance depends on the reasoning demands of the target task, the characteristics of the analysed artefacts, and the required outputs. Rather than identifying a universally superior model or prompting strategy, the results show that successful LLM adoption requires selecting models and prompts that are appropriate for the specific RE activity. Human expertise nevertheless remains essential wherever engineering judgement, factual grounding, and validation are required.

Beyond the individual empirical studies, this work contributes cumulative evidence on the practical readiness of current LLMs for RE. It demonstrates the value of evaluating LLMs across complementary RE activities rather than in isolation, providing a broader understanding of where these models are already effective and where important limitations remain. This cross-task perspective also offers practical guidance for RE/SE practitioners seeking to integrate LLMs into their engineering workflows.

Finally, the released replication package, including datasets, prompts, scripts, and experimental materials, supports transparency and reproducibility~\cite{supplementary_materials}. We hope this work encourages future empirical studies spanning multiple RE activities, enabling a more cumulative understanding of the role of LLMs throughout the RE lifecycle.
\backmatter

\bmhead{Supplementary information}
A replication package containing the datasets, prompts, scripts, and experimental materials is provided as supplementary material for peer review~\cite{supplementary_materials}.

\bmhead{Acknowledgements}
Study 1 was partially conducted within the MSc thesis of M. A. Mallya, supervised by J. D\k{a}browski~\cite{Mallya2025}. Study 2 was largely conducted during a research visit by J. D\k{a}browski to the Huawei Ireland Research Center in Dublin, Ireland. The results contribute to the Prompt Me project~\cite{Dabrowski2024}. This publication has emanated from research jointly funded by Taighde \'{E}ireann -- Research Ireland under Grant Number 13/RC/2094\_2 and co-funded by the European Union under the Systems, Methods, Context (SyMeCo) programme, Grant Agreement Number 101081459. Views and opinions expressed are, however, those of the author(s) only and do not necessarily reflect those of the European Union or the European Research Executive Agency. Neither the European Union nor the granting authority can be held responsible for them.

\section*{Declarations}
The authors declare that they have no known competing financial interests or personal relationships that could have appeared to influence the work reported in this paper.

%






\bibliography{sn-bibliography}

\end{document}